\documentclass[prb,twocolumn,preprintnumbers,superscriptaddress]{revtex4-2}

\usepackage{graphicx,epstopdf}% Include figure files
\usepackage{amssymb,amsmath}
\usepackage{dcolumn}% Align table columns on decimal point
\usepackage{bm}% bold math
\usepackage{color}
\usepackage{enumerate}
\usepackage[colorlinks=true,citecolor=blue]{hyperref}
\hypersetup{colorlinks=true,citecolor=blue,linkcolor=blue,urlcolor=blue}
\usepackage{soul}

\usepackage{tabularx} 
\usepackage[dvipsnames]{xcolor}

\begin{document}

\title{Theory of magnetotrion-polaritons in transition metal dichalcogenide monolayers}

\author{A. Kudlis}
\email{andrewkudlis@gmail.com}
\affiliation{Science Institute, University of Iceland, Dunhagi 3, IS-107, Reykjavik, Iceland}
\affiliation{Abrikosov Center for Theoretical Physics, Dolgoprudnyi, Moscow Region 141701, Russia}
\affiliation{Russian Quantum Center, Skolkovo, Moscow 121205, Russia}

\author{I. A. Aleksandrov}
\email{i.aleksandrov@spbu.ru}
\affiliation{Department of Physics, Saint Petersburg State University, Universitetskaya Naberezhnaya 7/9, Saint Petersburg 199034, Russia}
\affiliation{Ioffe Institute, Politekhnicheskaya street 26, Saint Petersburg 194021, Russia}

\author{K. Varga}
\email{kalman.varga@vanderbilt.edu}
\affiliation{Department of Physics and Astronomy, Vanderbilt University, Nashville, Tennessee 37235, USA}

\author{I. A. Shelykh}
\affiliation{Science Institute, University of Iceland, Dunhagi 3, IS-107, Reykjavik, Iceland}
\affiliation{Russian Quantum Center, Skolkovo, Moscow 121205, Russia}
\affiliation{Department of Physics, ITMO University, Saint Petersburg 197101, Russia}

\author{V. Shahnazaryan}
\email{vanikshahnazaryan@gmail.com}
\affiliation{Abrikosov Center for Theoretical Physics, Dolgoprudnyi, Moscow Region 141701, Russia}
\affiliation{Department of Physics, ITMO University, Saint Petersburg 197101, Russia}

\begin{abstract}
Magnetic field is a powerful tool for the manipulation of material's electronic and optical properties. In the domain of transition metal dichalcogenide monolayers, it allows one to unveil the spin, valley, and orbital properties of many-body excitonic complexes. Here we study theoretically the impact of normal-to-plane magnetic field on trions and trion-polaritons. We demonstrate that spin and orbital effects of a magnetic field give comparable contributions to the trion energies. Moreover, as magnetic field redistributes the free electron gas between two valleys in the conductance band, the trion-photon coupling becomes polarization and valley dependent. This results in an effective giant Zeeman splitting of trion-polaritons, in-line with the recent experimental observations.
\end{abstract}
%\date{\today}

\maketitle

\section{Introduction}

Monolayers of transition metal dichalcogenides (TMD) are representatives of the family of two-dimensional (2D) materials which reveal robust excitonic response. The corresponding bright exciton binding energies reach hundreds of meV \cite{Mak2010}, and  the presence of tightly bound excitons with large optical oscillator strength determine  the optical spectra of TMD monolayers up to room temperatures \cite{Wang2018}.
Moreover, in contrast to the case of conventional semiconductor quantum wells, excited excitonic states are clearly seen in the reflection spectrum \cite{chernikov2014exciton}.  They are characterized by non-Rydberg energy scaling resulting from the reduced screening of Coulomb interaction peculiar for a truly 2D system \cite{Rytova1967,Keldysh1979,berkelbach2013theory,huser2013dielectric}.

TMD monolayers have a hexagonal lattice structure with band edge minima at $\pm$K points of the Brillouin zone forming two non-equivalent valleys characterized by opposite spin polarizations \cite{kormanyos2015k}.
This gives rise to emergent spin-valley physics \cite{srivastava2015valley}, which in the domain of excitonics is characterized by a large variety of exciton species \cite{mostaani2017diffusion}, and determines the great potential of the considered materials for nanoscale device applications \cite{mueller2018exciton,anantharaman2021exciton,ciarrocchi2022excitonic}.

A direct consequence of tightly bound character of excitons in TMD monolayers is an ability to reach the regime of high density of excitons with $n\sim 10^{12}$--$10^{13}$~cm$^{-2}$ \cite{chernikov2015population,steinhoff2017exciton}, which is orders of magnitude larger than in conventional 2D systems based on GaAs quantum wells \cite{estrecho2019direct}.
In this regime the nonlinear optical effects associated with exciton-exciton Coulomb interactions become prominent \cite{shahnazaryan2017exciton,shahnazaryan2020tunable,bleu2020polariton,fey2020theory,erkensten2021exciton,cam2022symmetry}. Such nonlinear behavior is further enhanced in the presence of an optical microcavity, where the regime of strong light-matter coupling occurs and exciton-polaritons are formed~\cite{dufferwiel2015exciton}.

Excitonic and polaritonic properties of TMD monolayers are strongly affected by the presence of free electron gas. The unintentional doping routinely appears during the fabrication process, and the density of excess electrons can be controlled via external gating field \cite{chernikov2015electrical}, which thus becomes a control parameter for tuning of the exciton resonance. Moreover, the presence of free charge carriers results in a formation three-body charged excitonic complexes -- trions, characterized by a new emergent peak below the excitonic line in the optical spectrum 
\cite{mak2013tightly,ross2013electrical,singh2016trion,courtade2017charged}.
It should be noted that at high densities of free carriers the underlying physics of exciton-electron interaction becomes more complex, and a crossover from trions to repulsive (exciton) and attractive (trion) polarons occurs \cite{sidler2017fermi,efimkin2017many,tan2020interacting,efimkin2021electron}.
Yet, at low densities of itinerant electrons, the exciton-trion picture is sufficient for description of the optical response \cite{glazov2020optical}. 

Similar to excitons, trions are efficiently coupled to light. The regime of strong light-matter coupling in microcavities with doped TMD monolayers is well elaborated~\cite{dufferwiel2017valley,zhumagulov2022microscopic}. Interestingly, the nonlinear response associated with trion polaritons is larger as compared to excitons~\cite{emmanuele2020highly,kyriienko2020nonlinear}, and can be further enhanced by means of an external magnetic field~\cite{lyons2022giant}, which was also used to reveal the Rydberg series, and determine effective masses and $g$-factors of excitons \cite{stier2016exciton,stier2018magnetooptics,goryca2019revealing,have2019excitonic,liu2019magnetophotoluminescence,delhomme2019magneto}. However, the existing studies of the magnetic field impact on trions in TMD monolayers focus on a valley Zeeman splitting only \cite{macneill2015breaking,klein2021controlling} with orbital effects being typically neglected \cite{glazov2020optical}.
In this paper we thus develop a microscopic formalism where orbital and spin effects of a magnetic field are treated on equal footing, demonstrating that in the case of trions they give comparable contributions. We also demonstrate the vital role played by the mixing of different orbital channels.

%%%%%%%%%%%%%%%%%%%%

\section{Model}

\subsection{Trion states in the presence of a magnetic field}

\begin{figure}[t!]
    \centering
    \includegraphics[width=0.99\linewidth]{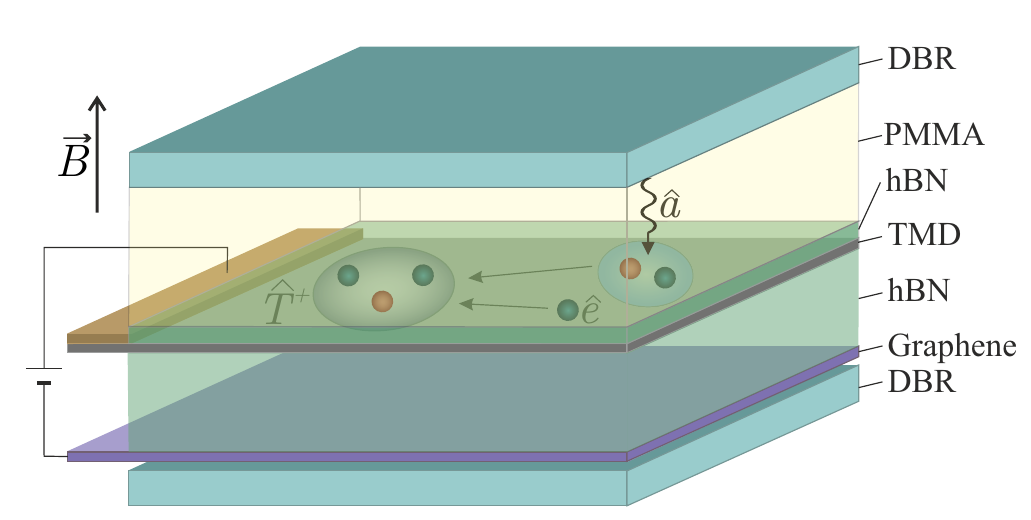}
    \caption{Sketch of the structure. A hBN-encapsulated TMD monolayer hosting excess electrons is embedded in an optical microcavity in the presence of a normal-to-plane uniform magnetic field. 
    The microcavity is filled with dielectric media represented by PMMA.
    A pair of electrodes consisting of a metallic gate and graphene layer is used to control the density of excess electrons. 
    A cavity photon creates an electron-hole pair, which is associated with a free electron forming a three-particle bound state (a trion). 
    The radiative recombination of a trion generates a cavity photon and a free electron. 
    In high-$Q$ optical cavity, a successive repetition of these processes leads to the onset of strong light-matter coupling regime and formation of hybrid states, trion-polaritons. }
    \label{fig:sketch}
\end{figure}

\begin{figure}[t]
    \centering
    \includegraphics[width=0.89\linewidth]{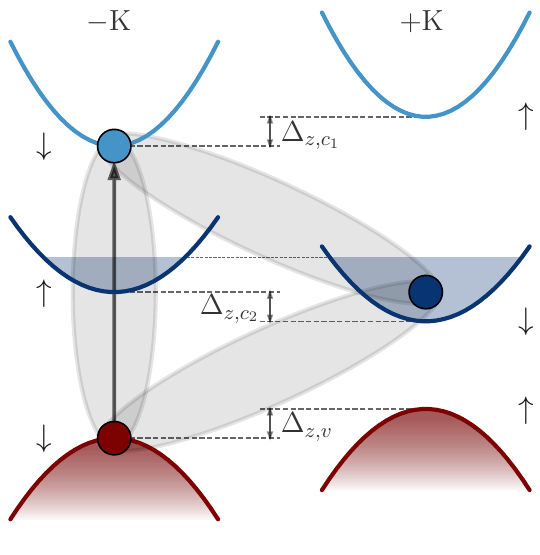}
    \caption{Band structure in the vicinity of $\pm$K points of Brillouin zone and optical excitonic transition (thick arrow) in WSe$_2$ monolayer in the presence of free electron gas.
    The arrows indicate the subband's spin projection. 
    The lowest-energy trion state in the -K valley is formed by an electron from the upper conduction subband and a hole from the upper valence subband, coupled with an excess electron, located in the +K valley of the lower conduction subband.}
    \label{fig:band}
\end{figure}

We consider a trion state in the presence of a normal-to-plane uniform magnetic field ${\bm   B}$ in monolayer WSe$_2$ placed inside a planar optical cavity represented by a pair of distributed Bragg reflectors (DBR), as shown in Fig.~\ref{fig:sketch}. 
The excess electrons are located in the lowest spin split subband in the vicinity of K~point of Brillouin zone, whereas an optical interband transition occurs to the upper subband (see Fig.~\ref{fig:band}).
This separation allows one to discard the phase space filling effects associated with Pauli blocking \cite{glazov2020optical}. 
The Hamiltonian of a three-particle system consisting of two electrons with coordinates ${\bm   r}_{1,2}$ and a hole with coordinate ${\bm   r}_h$ is given by  
\begin{align}
    \hat{H}_{\rm T}=\hat{H}_{\rm kin} +\hat{V}_{ eh}+\hat{V}_{ ee},  
\end{align}
where
\begin{align}
    \hat{H}_{\rm kin}&= \dfrac{\hat{{\bm   \pi}}_1^2}{2\mu_{e1}}
    +\dfrac{\hat{{\bm   \pi}}_2^2}{2\mu_{e2}}
    +\dfrac{\hat{{\bm   \pi}}_h^2}{2\mu_h},  \\
    \hat{V}_{ eh}&=-\sum\limits_{j=1,2}
    V(|{\bm   r}_j - {\bm   r}_h|),\\
    \hat{V}_{ ee}&= V(|{\bm   r}_1 - {\bm   r}_2|).
\end{align}
Here the momenta operators read $\hat{{\bm   \pi}}_j = -i \hbar {\bm   \nabla}_j - e {\bm   A} ({\bm   r}_j)$, $\hat{{\bm   \pi}}_h = -i \hbar {\bm   \nabla}_h + e {\bm   A} ({\bm   r}_h)$, $e$ is the positive unit charge,
and ${\bm   A}$ is the vector gauge potential, $\nabla \times {\bm   A} = {\bm   B}$. The parameters $\mu_{e1}$, $\mu_{e2}$, $\mu_h$ are the effective masses of upper and lower subband  electron, and a hole, respectively. 
The Coulomb interaction between charge carriers can be described  by means of the Rytova-Keldysh potential~\cite{Rytova1967,Keldysh1979}:
\begin{align}\label{eq:K-R_potential}
    V(r)= \frac{e^2}{4\pi  \varepsilon_0}
    \dfrac{\pi}{2 r_0}\left[H_0\left(\dfrac{\kappa r}{r_0}\right)-Y_0\left(\dfrac{\kappa r}{r_0}\right)\right],
\end{align}
where $\varepsilon_0$ is the static dielectric constant, $r_0$ is the effective screening length, $\kappa$ is the effective dielectric permittivity of surrounding media represented by a hexagonal boron nitride (hBN).
$H_0$ and $Y_0$  are the Struve function and Bessel  function of the second kind, respectively.

\begin{figure}[b!]
    \centering
    \includegraphics[width=1\linewidth]{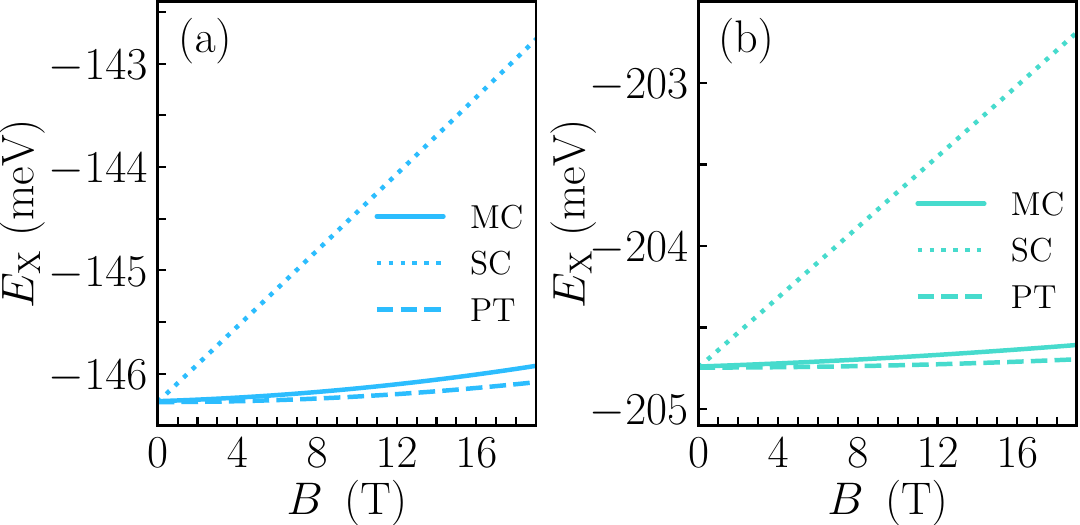}
    \caption{Exciton binding energy versus the magnetic field calculated for atomic monolayer (a)
    WSe$_2$, and (b) MoSe$_2$ by means of three different approaches.
    The dotted lines correspond to the single channel (SC) calculations within the SVM method, where one orbital channel $(m_e,m_h)=(0,0)$ is taken into account as discussed in Ref.~\cite{PhysRevB.97.195408}.
    The resulting diamagnetic shift scales linearly with the magnetic field and is essentially larger compared to the experimental measurements.
    The dashed lines correspond to the calculation within perturbation theory (PT), where the diamagnetic shift is estimated via $\Delta E_{\textup{dia}} \approx e^2\langle r^2 \rangle B^2/8\mu$, in accordance with the experimental results~\cite{stier2018magnetooptics,goryca2019revealing,have2019excitonic,liu2019magnetophotoluminescence,delhomme2019magneto}.
    The solid lines show the SVM calculations of the exciton energy with a multi-orbital basis incorporating the following orbital channels: $(0,0)$, $(1,-1)$, $(-1,1)$, $(-2,2)$,  and $(2,-2)$.}
    \label{fig:excitonbinding}
\end{figure}

We employ a variational approach for the calculation of the ground state energy and wave function of the bound state.
The numerical efficiency of the method strongly depends on the type of a trial function. 
The possible choice is the use of the anzats of fully correlated two-dimensional Gaussians \cite{varga2008solution,kidd2016binding}:
\begin{align}
&\!\!\Phi_{MS}^n({\bm   r})\!=\!\mathcal{A}\!\left\{\!\!\left[\prod\limits_{k=1}^3\!(x_k\!+\!i y_k)^{m_k^n}\!\right]\!   \right. \nonumber   \\
&\qquad \qquad \quad \qquad \qquad \times  \left.\exp{\!\!\left[\!-\dfrac{1}{2}\!\!\sum\limits_{i,j=1}^3\!\!\!  A^n_{ij}{\bm   r}_i{\bm   r}_j\!\right]\!}\chi_{SM_S}\right\},\!\!
\label{eq:Phi}
\end{align}
where $\mathcal{A}$ is the antisymmetrization operator, ${\bm   r}=({\bm   r}_1,{\bm   r}_2,{\bm   r}_3)$, the superindex $M=(m_1,m_2,m_3)$, $\chi_{SM_S}$ are linear combinations of single-particle spin functions corresponding to a given spin $S$ and spin projection $M_S$. Throughout the paper we assume ${\bm   r}_3\equiv{\bm   r}_h$. 

The basis functions~\eqref{eq:Phi} are used to construct a trial wave function as a linear combination $\Psi=\sum_i c_i\Phi_i$, where the superindex $i$ carries information about the orbital and spin quantum  numbers.
The functions $\Phi_i$ should be  optimized within the procedure  of random parameter modification in  order  to  minimize the ground state energy. 
Given that these functions are not orthogonal to each other, one has to solve a generalized eigenvalue problem:
\begin{align}
    \label{eq:eval}
    &\sum\limits_{j}\left(H_{ij}-E O_{ij}\right)c_j=0, \\
    &H_{ij}=\langle\Phi_i|\hat{H}_{\rm T}|\Phi_j\rangle, \quad  O_{ij}=\langle\Phi_i|\Phi_j\rangle, 
\end{align}
where as the output we obtain $E_1$, $E_2$, $E_3$, $\dots$ as variational upper bounds of energies of respective states and corresponding eigenvectors. In what follows we denote the trion   ground state by $\Psi_{\rm T} ({\bm r}_1, {\bm  r}_2, {\bm  r}_3)$. We perform a similar procedure for the exciton problem, where the ground-state wave function will be denoted by $\Psi_{\rm X} ({\bm r}_1, {\bm  r}_2)$.

The parameters of matrix $A_{ij}^n$ can be found by means of the stochastic variational method (SVM), which has proven its high efficiency in 
typical eigenvalue problems for few-body systems~\cite{VARGA1997157,Suzuki2014-mg,PhysRevC.52.2885,PhysRevB.63.205308}. 
Detailed description of the numerical procedure is presented in Refs.~\cite{varga2008solution,kidd2016binding,PhysRevB.97.195408,PhysRevB.101.235435}. 
Here we use the \textit{svm-2d} package~\cite{varga2008solution}, in which we recalculate the matrix elements of the interparticle interaction with the potential~\eqref{eq:K-R_potential} by Gauss–Legendre quadrature.
We first make sure that in the absence of the magnetic field, the resulting trion binding energy matches with the previous reports~\cite{PhysRevB.101.235435}. 
As we only consider the ground state, we set for an exciton $S=0$, $M_S=0$, and for a trion $S=1/2$, $M_S=1/2$. 

\begin{table}[]
    \centering
     \setlength{\tabcolsep}{5.6pt}
    \caption{Characteristics of the materials considered in this work. Effective masses are taken from Ref.~\cite{kormanyos2015k}, and screening lengths from Ref.~\cite{goryca2019revealing}. Here $E_{\rm b}^{\rm X}$ ($E_{\rm b}^{\rm T}$) is the calculated exciton (trion) binding energy in the absence of the magnetic field.} 
\renewcommand{\arraystretch}{1.2} % Default value: 1
    \begin{tabular}{cccccccc} 
      \hline
      \hline
        parameter & $\mu_{e1}$ & $\mu_{e2}$ & $\mu_h$ & $\kappa$ & $r_0$ & $|E_{\rm X}|$ & $|E_{\rm T}^{\rm b}|$  \\ 
        unit & $m_0$ & $m_0$ & $m_0$ &  & nm & meV & meV \\
              \hline
        WSe$_2$ & 0.28 & 0.39 & 0.36 & 4.4 & 4.5 & $146.3$  & $20.8$  \\
        WS$_2$ & 0.26 & 0.35 & 0.36 & 4.4 & 3.4 & $162.8$ &  $23.5$  \\ 
        MoSe$_2$ & 0.56 & 0.49 & 0.59 & 4.4 & 3.9 & $204.7$ & $16.9$  \\ 
        MoS$_2$ & 0.46 & 0.43 & 0.54 & 4.4 & 3.4 & $207.8$ & $17.5$  \\
      \hline
      \hline
        \end{tabular}
    \label{tab:materials}
\end{table}

\subsection{Trion-photon coupling}

We consider a direct interband optical transition within a two-band approximation. We consider the case of the normal incidence, so that the photon wave vector ${\bm   q}_{\rm ph}=0$. The corresponding Hamiltonian can be represented as
\begin{align}
    \hat{\mathcal{H}}_{\rm rad} 
    =\Omega\sum_{{\bm   k}_e,{\bm   k}_h} 
    \left( \hat{a}_{{\bm   k}_e}^\dag \hat{b}_{{\bm   k}_h}^\dag
    + \hat{b}_{{\bm   k}_h} \hat{a}_{{\bm   k}_e} \right) \delta_{ {\bm   k}_e + {\bm   k}_h , 0} 
\end{align}
with
\begin{equation}
    \Omega = d_{cv} \sqrt{\frac{\hbar\omega_{\rm C}}  {2\varepsilon\varepsilon_0 L_{\rm C} S} },    
\end{equation}
where $\omega_{\rm C}$ is the cavity eigenfrequency, 
$\varepsilon$ is the dielectric constant of a media typically represented by an organic polymer, such as polymethyl methacrylate (PMMA) \cite{lundt2016room,gu2019room,anton2021bosonic}. Then $L_{\rm C}$ and $S$ are the cavity length and sample area, respectively, $d_{cv}$ is the dipole matrix element of the interband transition, $\hat{a}_{{\bm   k}_e}^\dag$, $\hat{b}_{{\bm   k}_h}^\dag$ are the creation operators for electrons and holes, respectively. The trion state can be expressed in the form
\begin{align}
    |T\rangle = \sum_{{\bm   k}_1,{\bm   k}_2,{\bm   k}_h} F_{\rm T} ({\bm   k}_1,{\bm   k}_2,{\bm   k}_h) \hat{a}_{{\bm   k}_1}^\dag
    \hat{\tilde{a}}_{{\bm   k}_2}^\dag
    \hat{b}_{{\bm   k}_h}^\dag 
    | \text{\o} \rangle ,
\end{align}
where $\hat{\tilde{a}}_{{\bm   k}_2}^\dag$ corresponds to excess electrons, and $F_{\rm T} ({\bm   k}_1,{\bm   k}_2,{\bm   k}_h)$ is the trion wave function in the momentum space. Accordingly, the excess electron state is $|e\rangle = \hat{\tilde{a}}_{{\bm   k}}^\dag
| \text{\o} \rangle$.

\begin{figure}[t!]
    \centering
    \includegraphics[width=1\linewidth]{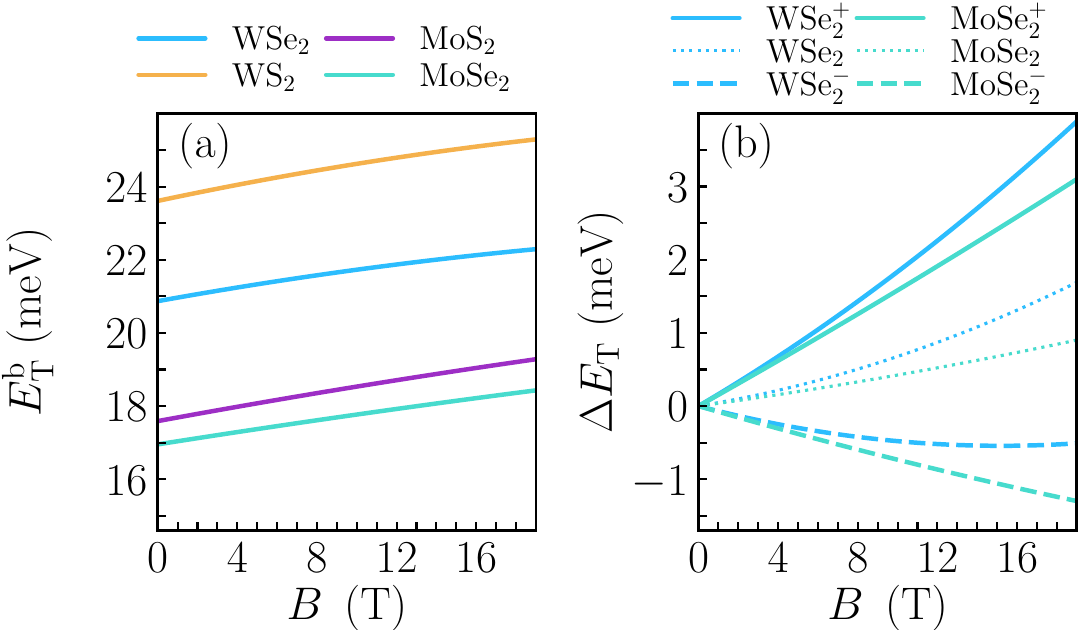}
    \caption{(a)~Trion binding energy versus the magnetic field, calculated using the expression~\eqref{eqn:trion_b_e} for various TMD monolayers. (b)~Trion resonance energy as a function of the magnetic field for WSe$_2$ and MoSe$_2$ atomic monolayers. 
    The dotted lines correspond to the diamagnetic shift. 
    The solid and dashed lines demonstrate the total shift accounting for the Zeeman splitting for the two spin orientations.
    }
    \label{fig:trionbinding}
\end{figure}

The trion-photon coupling Hamiltonian then reads
\begin{align}
    \hat{H}_{\rm TC} = \sum_{\bm   k} \left( \hat{T}_{\bm   k}^\dag \langle T| \hat{\mathcal{H}}_{\rm rad} | e \rangle \hat{\tilde{a}}_{{\bm   k}} \hat{c} 
    +\hat{c}^\dag \hat{\tilde{a}}_{{\bm   k}}^\dag \langle e| \hat{\mathcal{H}}_{\rm rad} | T \rangle  \hat{T}_{\bm   k}
    \right).
\end{align}
The coupling rate is then evaluated via
\begin{align}
    \langle T| \hat{\mathcal{H}}_{\rm rad} | e \rangle  = &\Omega \sum_{{\bm   k}_1,{\bm   k}_2,{\bm   k}_h} 
    \sum_{{\bm   k}'_e,{\bm   k}'_h}
    F^*_{\rm T} ({\bm   k}_1,{\bm   k}_2,{\bm   k}_h) \delta_{ {\bm   k}'_e + {\bm   k}'_h , 0} \notag \\
    &\langle \text{\o} |
    \hat{b}_{{\bm   k}_h}
    \hat{\tilde{a}}_{{\bm   k}_2}  
    \hat{a}_{{\bm   k}_1}  
    \left( \hat{a}_{{\bm   k}'_e}^\dag \hat{b}_{{\bm   k}'_h}^\dag
    + \hat{b}_{{\bm   k}'_h} \hat{a}_{{\bm   k}'_e} \right) 
    \hat{\tilde{a}}_{\bm   k}^\dag  | \text{\o} \rangle \notag \\
    {}= &\Omega
    \sum_{{\bm   k}_h} 
    F^*_{\rm T} (-{\bm   k}_h,{\bm   k},{\bm   k}_h)
    =\Omega I_{\rm T}({\bm   k}).
\end{align}
In the presence of a magnetic field, the center-of-mass dynamics of trions cannot be straightforwardly decoupled from its internal dynamics, and one obtains
\begin{align}
    \label{eq:IT}
    &I_{\rm T} ({\bm   k}) = \sum_{{\bm   k}_h}  F^*_{\rm T} (-{\bm   k}_h, {\bm   k}, {\bm   k}_h) \notag \\
    &= \frac{S}{(2\pi)^2}  \frac{1}{ S^{\frac{3}{2}} } \int {\rm d}^2 {\bm   k}_h {\rm d}^2 {\bm   r}_1 {\rm d}^2 {\bm   r}_2 {\rm d}^2 {\bm   r}_h 
    e^{-i{\bm   k}_h {\bm   r}_1}
    e^{i{\bm   k} {\bm   r}_2}
    e^{i{\bm   k}_h {\bm   r}_h}
    \notag \\
    &\times  \Psi_{\rm T} ({\bm   r}_1, {\bm   r}_2, {\bm   r}_h) = \frac{1}{ \sqrt{S} } \int  {\rm d}^2 {\bm   r}_2 {\rm d}^2 {\bm   r}_h 
    e^{i{\bm   k} {\bm   r}_2}
    \Psi_{\rm T} ({\bm   r}_h, {\bm   r}_2, {\bm   r}_h) . 
\end{align}

For the case of moderate dopings $|{\bm   k}| \sim |{\bm   k}_F| \ll a_{\rm tr}^{-1}$,
where $a_{\rm tr} = \langle \Psi_{\rm T}  |r| \Psi_{\rm T} \rangle$ is the effective Bohr radius of trion, and ${\bm   k}_F \propto \sqrt{n_e}$
is the Fermi wave vector defined by excess electron density $n_e$.
Hence, one can assume $e^{i {\bm   k} {\bm   r}} \approx 1$, so that  $I_{\rm T} ({\bm   k}) \approx I_{\rm T}= \mathrm{const}$~\cite{glazov2020optical}.
The total electron density is the sum of the densities in two valleys, $n_e = n_{e,+{\rm K}} + n_{e,-{\rm K}}$, where 
\begin{align}
    \label{eq:npm}
    n_{e,\pm {\rm K}} = \frac{k_{\rm B} T}{2E_{\rm F}} \ln \left[\exp\left( \frac{E_{\rm F} \pm \Delta E_{{\rm Z},c2}/2 }{k_{\rm B} T} \right) +1 \right],
\end{align}
with $E_{\rm F} = \pi \hbar^2 n_e / (2 \mu_{e2})$  being the Fermi energy in the absence of the magnetic field, $k_{\rm B}$ denoting Boltzmann constant, and $T$ standing for the temperature. The magnetic field induces a redistribution of free charges between the two valleys, determined by the Zeeman splitting of the lower conduction subband $\Delta E_{{\rm Z},c2} = g_{c2} \mu_{\rm B} B$, where $\mu_{\rm B}$ is the Bohr magneton, 
$g_{c2}$ is the Lande factor of the lower conduction subband. 
In the limit $E_{\rm F} > \Delta E_{{\rm Z},c2} \gg k_{\rm B} T$, the expression~\eqref{eq:npm} reduces to $n_{e,\pm {\rm K}} = \left[1 \pm \Delta E_{{\rm Z},c2} / (2E_{\rm F})  \right] n_e/2$.

We further introduce quasi-bosonic operators~\cite{emmanuele2020highly}
\begin{align}
    \hat{\mathcal{T}}^\dag_{\pm} = \frac{1}{\sqrt{N_{e, {\rm \pm K} }}} \sum_{\bm   k} \hat{T}_{\bm   k}^\dag \hat{\tilde{a}}_{{\bm   k}} ,
\end{align}
where $N_{e, {\rm \pm K} } =n_{e, {\rm \pm K} } S$ is the total number of excess electrons in each valley. Then we find
\begin{align}
    \hat{H}_{\rm TC}^{\pm} =  \Omega I_T \sqrt{N_{e, {\rm \pm K} }} \left(\hat{\mathcal{T}}^\dag_{ \pm} \hat{c} 
    +\hat{c}^\dag \hat{\mathcal{T}}_{\pm}    \right) 
    = \Omega_{\rm T}^{\pm}  \left(\hat{\mathcal{T}}^\dag_{\pm} \hat{c} 
    +\hat{c}^\dag \hat{\mathcal{T}}_{\pm}    \right),
\end{align}
where 
\begin{align}
    \label{eq:OmegaT}
    \Omega_{\rm T}^{\pm} =
    \sqrt{n_{e,\pm {\rm K}} } \Omega  \int  {\rm d}^2 {\bm   r}_2 {\rm d}^2 {\bm   r}_h 
    \Psi_{\rm T} ({\bm   r}_h, {\bm   r}_2, {\bm   r}_h) .
\end{align}
In the absence of the magnetic field, the trion wave function can be factorized as 
$\Psi_{\rm T} ({\bm r}_1, {\bm r}_2, {\bm r}_h)|_{{\bm B} =0} =  \psi_{\rm T} ({\bm r}_1 - {\bm r}_h, {\bm r}_2 - {\bm r}_h)/\sqrt{S}$, where $\psi_{\rm T}$ is the wave function of relative dynamics, and the center-of-mass dynamics is represented by a plane wave with a zero wave-vector.
Then the trion-photon coupling is simplified:
\begin{align}
    \label{eq:OmegaT0}
    \Omega_{\rm T}^{\pm} |_{{\bm B} =0}  =
    \sqrt{n_{e,\pm {\rm K}} } \Omega \sqrt{S} \int  {\rm d}^2 {\bm   r }  
    \psi_{\rm T} (0, {\bm   r}) .
\end{align}

Analogously, the exciton-photon coupling reads~\cite{glazov2020optical}
\begin{align}
    \label{eq:OmegaX}
    \Omega_{\rm X} = \Omega \int {\rm d}^2 {\bm   r} \Psi_{\rm X} ({\bm   r}, {\bm   r}).
\end{align}
Note that similarly to the case of trions, the dynamics of center of mass cannot be decoupled from the internal dynamics if the magnetic field is present.
In the absence of the magnetic field, one has $\Psi_{\rm X} ({\bm r}_e, {\bm r}_h) =  \psi_{\rm X} (|{\bm   r}_e - {\bm   r}_h|) / \sqrt{S}$, where $\psi_{\rm X}$ is the wave function of relative dynamics involved in the well-known expression $\Omega_{\rm X} = \Omega \sqrt{S} \psi_{\rm X} (0)$~\cite{claudio1995optical}. 

Given the large energy separation (above 30 meV) between the exciton and trion optical transitions~\cite{courtade2017charged}, we consider a strong coupling of trions and cavity modes \textit{only}, i.e., we disregard the exciton resonance. The corresponding Hamiltonian reads
\begin{align}
    \label{eq:HamTC}
    \hat{H}^{\pm}  = \begin{pmatrix}
            \hbar \omega_{\rm C}^{\pm} &  \Omega_{\rm T}^{\pm} \\
            \Omega_{\rm T}^{\pm} & E_{\rm T}^{\pm} 
        \end{pmatrix} ,
\end{align}
where the superscript $\pm$ denotes circular polarization of the cavity mode and the trion valley index. The eigenmodes of this Hamiltonian correspond to the upper and lower trion-polaritons.

%%%%%%%%%%%%%%%%%%%%

\begin{figure}[t]
    \centering
  \includegraphics[width=1\linewidth]{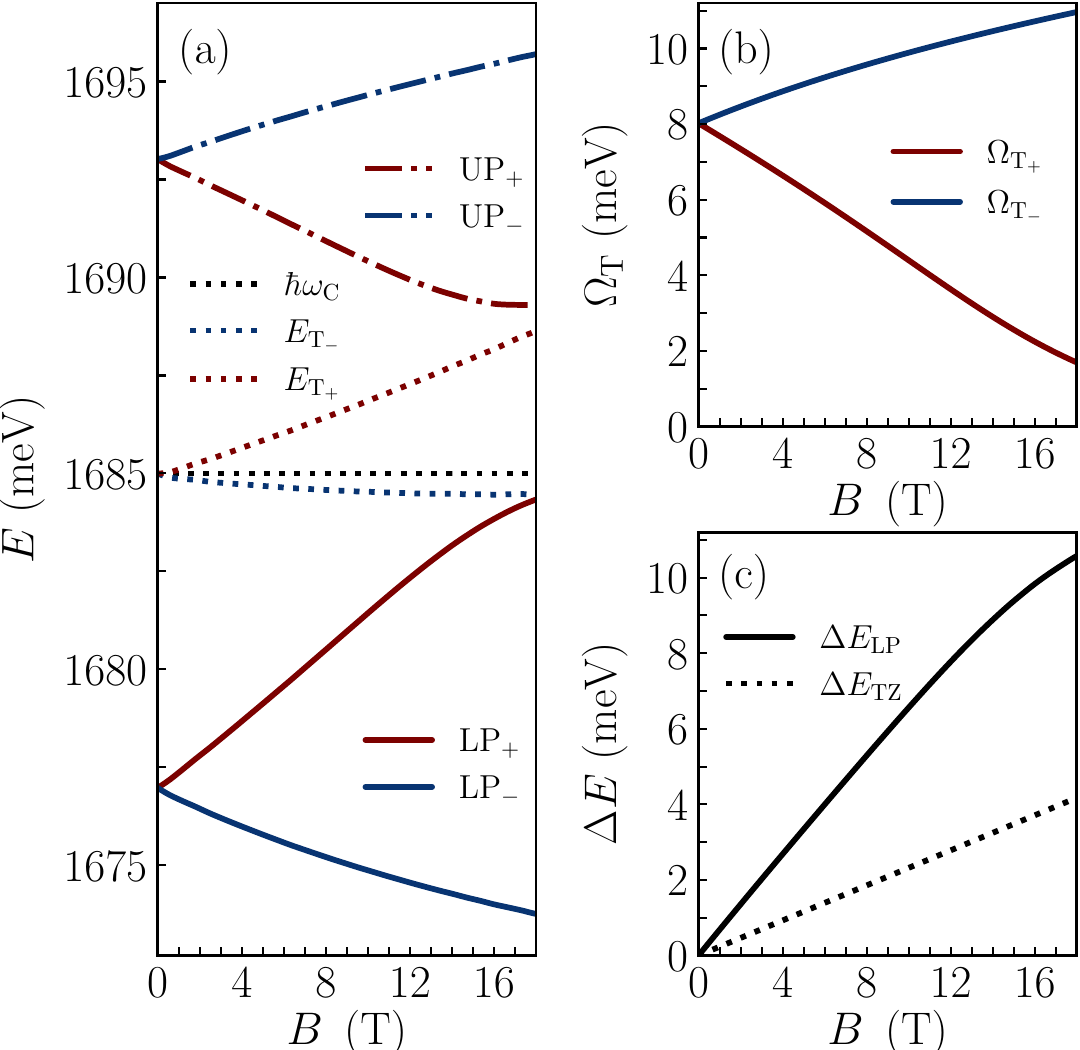}
    \caption{(a)~Energy of trion-polariton branches versus the magnetic field for atomic monolayer   WSe$_2$. 
    Red [blue] curves correspond $\sigma^{+[-]}$ circular polarizations.
    The increase of the magnetic field leads to a collapse of Rabi splitting for $\sigma^{+}$-polarized trions, and the respective increase of Rabi splitting for $\sigma^{-}$-polarized trions.
    (b)~Trion-photon coupling energy versus the magnetic field calculated via Eq.~\eqref{eq:OmegaT}.
    (c)~Zeeman splitting of bare trions (dashed line) and trion-polaritons (solid line).  
    The reduction [enhancement] of Rabi splitting with increasing magnetic field for $\sigma^{+[-]}$-polarized results in a large increase of Zeeman splitting of trion-polaritons.
    }
    \label{fig:trionRabi}
\end{figure}

%
%%%%%%%%%%%%%%%%%%%%
\section{Results and discussion}

\subsection{Trion states in the presence of a magnetic field}

We start with addressing the exciton binding energy in the absence of excess charge carriers. 
The exciton problem in a weak magnetic field is well studied~\cite{Gorkov1967}, including the case of TMD monolayers~\cite{stier2018magnetooptics,goryca2019revealing,have2019excitonic,liu2019magnetophotoluminescence,delhomme2019magneto}. The magnetic field results in a diamagnetic energy shift, which in the lowest order of perturbation theory amounts to $\Delta E_{\textup{dia}} \approx e^2\langle r^2 \rangle B^2/8\mu$,
where $\mu = \mu_{e1}\mu_h/ (\mu_{e1}+\mu_h)$ is the exciton reduced mass
and $\langle r^2 \rangle = \langle \psi_{\rm X} |r^2| \psi_{\rm X} \rangle$ is calculated in the absence of the magnetic field.
To treat the problem more rigorously, we invoke the SVM method. We employ two sets of basis functions.
The first one includes a single orbital channel --- $(m_e,m_h)=(0,0)$, as in the case considered in Ref.~\cite{PhysRevB.97.195408}.
The results of calculation are shown in Fig.~\ref{fig:excitonbinding}(a), (b) for monolayers WSe$_2$, and MoSe$_2$, respectively.
The material parameters are given in Table~\ref{tab:materials}.
While in the absence of the magnetic field, the exciton binding energy is well reproduced, this approach demonstrates a linear (instead of quadratic) scaling with respect to the magnetic field, and severely overestimates the diamagnetic shift.
On the contrary, by using the multi-orbital basis with the orbital channels $(0,0)$, $(1,-1)$, $(-1,1)$, $(-2,2)$, and $(2,-2)$, one obtains the results which agree well with the experimental data and with the simple perturbative estimate discussed above.

Let us numerically analyze the generalized eigenvalue problem~\eqref{eq:eval} for the trion states. We adopt the multi-orbital basis with all possible combinations $-2 \leq m_i \leq 2$ such that $\sum_i m_i=0$.
The results of the calculations are summarized in Fig. \ref{fig:trionbinding}. 
In the left graph, we display the trion binding energy as a function of the magnetic field for various TMD monolayers. 
First of all, we note that the value of the trion binding energy in monolayer WSe$_2$ is above 20~meV in the absence of the magnetic field. Together with the exciton binding energy of $|E_{\rm X}| = 146.3$~meV, this provides essentially better agreement with experimental data than the results reported in Ref.~\cite{courtade2017charged}. The larger value of the trion binding energy was obtained due to different effective masses of electrons in the spin subbands. The presence of a magnetic field leads to the increase of the trion binding energy defined as
\begin{equation}\label{eqn:trion_b_e}
    E_{\rm T}^{\rm b} = |E_{\rm T}|  - |E_{\rm X}| + \hbar \omega_{\rm e} / 2,
\end{equation}
where $E_{\rm T}$ is the ground-state eigenvalue of Eq.~\eqref{eq:eval},
$\omega_{\rm e} = e |{\bm B}| / \mu_{e1}$ is the cyclotron frequency of free electrons.

In Fig.~\ref{fig:trionbinding}(b) the trion resonance energy shift versus magnetic field for monolayers of WSe$_2$ and MoSe$_2$ is shown.
The pure diamagnetic shift defined as $\Delta E_{\rm T}^{\rm dia} ({\bm B}) = |E_{\rm T} ({\bm B}) - E_{\rm T}({\bm B}=0)|$ is depicted by the dotted lines.
We account for the magnetic-field induced Zeeman splitting of trion energy $\Delta E_{\rm T Z} = g_{\rm X} \mu_{\rm B} B$ \cite{glazov2020optical}, where $g_{\rm X} = g_c - g_v$ is the exciton Lande factor defined via the Lande factors of upper conduction subband $g_c$, and valence band $g_v$.
The overall shift of trion resonances for the two spin orientations $\Delta E_{\rm T}^{\pm} (B) = \Delta E_{\rm T}^{\rm dia} (B) \pm \Delta E_{\rm TZ}/2 $ is displayed in Fig.~\ref{fig:trionbinding}(b) by solid and dashed lines.
Our direct calculations indicate that at the scale of 10~T of the applied magnetic field, the diamagnetic shift of the trion energy is comparable with the energy of Zeeman splitting.
%

%%%%%%%%%%%%%%%%%%

\subsection{Trion-photon coupling}

We now consider the regime of strong light-matter coupling of trions with a near-resonant cavity eigenmode characterized by Eq.~\eqref{eq:HamTC}. 
The negative trion resonance in hBN-encapsulated WSe$_2$ in the absence of the magnetic field and at low density of free carriers corresponds to $E_{\rm T}^0 =1685$~meV~\cite{courtade2017charged}.
We use this value to choose the resonance energy of a cavity mode according to $\hbar \omega_{\rm C} = E_{\rm T}^0$.
The dielectric constant of the media is set $\varepsilon=3$, typical for PMMA~\cite{na2006electronic}, and the cavity length is found from the resonance condition in the form $L_{\rm C} = 2\pi c  / (\sqrt{\varepsilon} \omega_{\rm C})$, where $c$ is the speed of light.
We fix the density of excess electrons at $n_e = 5 \times 10^{11}$~cm$^{-2}$, and also set $T=4$~K.
The dipole matrix element of the interband transition is chosen as $d_{cv} = 18$~Deb, resulting in $\Omega_{\rm T} \approx 8$~meV in the absence of the magnetic field, typical for the trion-photon coupling~\cite{emmanuele2020highly,lyons2022giant,khestanova2024electrostatic}. 

The polarization-resolved magnetic field dependence of the polariton energies for monolayer WSe$_2$ is shown in Fig.~\ref{fig:trionRabi}(a).
The $\sigma^+$ and $\sigma^-$ polarized branches demonstrate the opposite behavior with increasing magnetic field: the Rabi splitting of the former reduces, while for the latter it increases. The reason is as follows. The magnetic field results in an imbalance of excess electrons, where a major part of them is shifted into -K valley, see Eq.~\eqref{eq:npm}. As the trion photon-coupling rate scales as $\Omega_{\rm T} \propto \sqrt{n_e}$, this gives rise to the reduction of trion-photon coupling for $\sigma^+$ polarization, and enhancement for $\sigma^-$ polarization, as shown in Fig.~\ref{fig:trionRabi}(b). 
We note here that generally the trion-photon coupling $\Omega_{\rm T}$ depends on the magnetic field via the factor $I_{\rm T}$ as well. However, our calculation demonstrated that this dependence is minor.

The valley-dependent nature of the trion-photon coupling $\Omega_{\rm T}$ in the presence of a magnetic field results in drastic modulation of the trion Zeeman splitting. 
As the lower trion-polariton energies of the two polarizations shift in opposite directions with increasing magnetic field, in the strong-coupling regime, the effective Zeeman splitting essentially increases, as shown in Fig.~\ref{fig:trionRabi}(c).  This result agrees well with a recent experimental observation of giant Zeeman splitting of trion-polaritons in MoSe$_2$-based structure~\cite{lyons2022giant}.

%%%%%%%%%%%%%%%%%%%%
\section{Conclusion}

We developed a microscopic formalism for the description of magnetotrions and magnetotrion-polaritons in TMD monolayers. The spin and orbital effects of the magnetic field were treated on the same footing, and it was demonstrated that they give comparable contributions to the trion energy. We also showed that if a trion resonance is coupled to an optical cavity mode, the Zeeman splitting in the conductance band leads to a strong valley and polarization dependence of the trion-polariton energies, ultimately resulting in an effective giant Zeeman splitting for trion-polaritons.

\section*{Acknowledgments}
The research is supported by the Ministry of Science and Higher Education of the Russian Federation (Goszadaniye) Project No. FSMG-2023-0011.
V.S. acknowledges the support of ‘Basis’ Foundation (Project No. 22-1-3-43-1). The work of A.K. is supported by the Icelandic Research Fund (Rannis, Grant No. 2410550-051). 
The work of K.V. has been supported by the National Science Foundation (NSF) under Grant No. 2217759.


\begin{thebibliography}{61}%
\makeatletter
\providecommand \@ifxundefined [1]{%
 \@ifx{#1\undefined}
}%
\providecommand \@ifnum [1]{%
 \ifnum #1\expandafter \@firstoftwo
 \else \expandafter \@secondoftwo
 \fi
}%
\providecommand \@ifx [1]{%
 \ifx #1\expandafter \@firstoftwo
 \else \expandafter \@secondoftwo
 \fi
}%
\providecommand \natexlab [1]{#1}%
\providecommand \enquote  [1]{``#1''}%
\providecommand \bibnamefont  [1]{#1}%
\providecommand \bibfnamefont [1]{#1}%
\providecommand \citenamefont [1]{#1}%
\providecommand \href@noop [0]{\@secondoftwo}%
\providecommand \href [0]{\begingroup \@sanitize@url \@href}%
\providecommand \@href[1]{\@@startlink{#1}\@@href}%
\providecommand \@@href[1]{\endgroup#1\@@endlink}%
\providecommand \@sanitize@url [0]{\catcode `\\12\catcode `\$12\catcode
  `\&12\catcode `\#12\catcode `\^12\catcode `\_12\catcode `\%12\relax}%
\providecommand \@@startlink[1]{}%
\providecommand \@@endlink[0]{}%
\providecommand \url  [0]{\begingroup\@sanitize@url \@url }%
\providecommand \@url [1]{\endgroup\@href {#1}{\urlprefix }}%
\providecommand \urlprefix  [0]{URL }%
\providecommand \Eprint [0]{\href }%
\providecommand \doibase [0]{https://doi.org/}%
\providecommand \selectlanguage [0]{\@gobble}%
\providecommand \bibinfo  [0]{\@secondoftwo}%
\providecommand \bibfield  [0]{\@secondoftwo}%
\providecommand \translation [1]{[#1]}%
\providecommand \BibitemOpen [0]{}%
\providecommand \bibitemStop [0]{}%
\providecommand \bibitemNoStop [0]{.\EOS\space}%
\providecommand \EOS [0]{\spacefactor3000\relax}%
\providecommand \BibitemShut  [1]{\csname bibitem#1\endcsname}%
\let\auto@bib@innerbib\@empty
%</preamble>
\bibitem [{\citenamefont {Mak}\ \emph {et~al.}(2010)\citenamefont {Mak},
  \citenamefont {Lee}, \citenamefont {Hone}, \citenamefont {Shan},\ and\
  \citenamefont {Heinz}}]{Mak2010}%
  \BibitemOpen
  \bibfield  {author} {\bibinfo {author} {\bibfnamefont {K.~F.}\ \bibnamefont
  {Mak}}, \bibinfo {author} {\bibfnamefont {C.}~\bibnamefont {Lee}}, \bibinfo
  {author} {\bibfnamefont {J.}~\bibnamefont {Hone}}, \bibinfo {author}
  {\bibfnamefont {J.}~\bibnamefont {Shan}},\ and\ \bibinfo {author}
  {\bibfnamefont {T.~F.}\ \bibnamefont {Heinz}},\ }\bibfield  {title} {\bibinfo
  {title} {Atomically thin ${\mathrm{mos}}_{2}$: A new direct-gap
  semiconductor},\ }\href {https://doi.org/10.1103/PhysRevLett.105.136805}
  {\bibfield  {journal} {\bibinfo  {journal} {Phys. Rev. Lett.}\ }\textbf
  {\bibinfo {volume} {105}},\ \bibinfo {pages} {136805} (\bibinfo {year}
  {2010})}\BibitemShut {NoStop}%
\bibitem [{\citenamefont {Wang}\ \emph {et~al.}(2018)\citenamefont {Wang},
  \citenamefont {Chernikov}, \citenamefont {Glazov}, \citenamefont {Heinz},
  \citenamefont {Marie}, \citenamefont {Amand},\ and\ \citenamefont
  {Urbaszek}}]{Wang2018}%
  \BibitemOpen
  \bibfield  {author} {\bibinfo {author} {\bibfnamefont {G.}~\bibnamefont
  {Wang}}, \bibinfo {author} {\bibfnamefont {A.}~\bibnamefont {Chernikov}},
  \bibinfo {author} {\bibfnamefont {M.~M.}\ \bibnamefont {Glazov}}, \bibinfo
  {author} {\bibfnamefont {T.~F.}\ \bibnamefont {Heinz}}, \bibinfo {author}
  {\bibfnamefont {X.}~\bibnamefont {Marie}}, \bibinfo {author} {\bibfnamefont
  {T.}~\bibnamefont {Amand}},\ and\ \bibinfo {author} {\bibfnamefont
  {B.}~\bibnamefont {Urbaszek}},\ }\bibfield  {title} {\bibinfo {title}
  {Colloquium: Excitons in atomically thin transition metal dichalcogenides},\
  }\href {https://doi.org/10.1103/RevModPhys.90.021001} {\bibfield  {journal}
  {\bibinfo  {journal} {Rev. Mod. Phys.}\ }\textbf {\bibinfo {volume} {90}},\
  \bibinfo {pages} {021001} (\bibinfo {year} {2018})}\BibitemShut {NoStop}%
\bibitem [{\citenamefont {Chernikov}\ \emph {et~al.}(2014)\citenamefont
  {Chernikov}, \citenamefont {Berkelbach}, \citenamefont {Hill}, \citenamefont
  {Rigosi}, \citenamefont {Li}, \citenamefont {Aslan}, \citenamefont
  {Reichman}, \citenamefont {Hybertsen},\ and\ \citenamefont
  {Heinz}}]{chernikov2014exciton}%
  \BibitemOpen
  \bibfield  {author} {\bibinfo {author} {\bibfnamefont {A.}~\bibnamefont
  {Chernikov}}, \bibinfo {author} {\bibfnamefont {T.~C.}\ \bibnamefont
  {Berkelbach}}, \bibinfo {author} {\bibfnamefont {H.~M.}\ \bibnamefont
  {Hill}}, \bibinfo {author} {\bibfnamefont {A.}~\bibnamefont {Rigosi}},
  \bibinfo {author} {\bibfnamefont {Y.}~\bibnamefont {Li}}, \bibinfo {author}
  {\bibfnamefont {B.}~\bibnamefont {Aslan}}, \bibinfo {author} {\bibfnamefont
  {D.~R.}\ \bibnamefont {Reichman}}, \bibinfo {author} {\bibfnamefont {M.~S.}\
  \bibnamefont {Hybertsen}},\ and\ \bibinfo {author} {\bibfnamefont {T.~F.}\
  \bibnamefont {Heinz}},\ }\bibfield  {title} {\bibinfo {title} {Exciton
  binding energy and nonhydrogenic rydberg series in monolayer ws 2},\ }\href
  {https://journals.aps.org/prl/abstract/10.1103/PhysRevLett.113.076802}
  {\bibfield  {journal} {\bibinfo  {journal} {Physical review letters}\
  }\textbf {\bibinfo {volume} {113}},\ \bibinfo {pages} {076802} (\bibinfo
  {year} {2014})}\BibitemShut {NoStop}%
\bibitem [{\citenamefont {Rytova}(1967)}]{Rytova1967}%
  \BibitemOpen
  \bibfield  {author} {\bibinfo {author} {\bibfnamefont {N.~S.}\ \bibnamefont
  {Rytova}},\ }\bibfield  {title} {\bibinfo {title} {The screened potential of
  a point charge in a thin film},\ }\href
  {http://vmu.phys.msu.ru/en/abstract/1967/3/1967-3-030/} {\bibfield  {journal}
  {\bibinfo  {journal} {Moscow University Physics Bulletin}\ }\textbf {\bibinfo
  {volume} {22}},\ \bibinfo {pages} {18} (\bibinfo {year} {1967})}\BibitemShut
  {NoStop}%
\bibitem [{\citenamefont {Keldysh}(1979)}]{Keldysh1979}%
  \BibitemOpen
  \bibfield  {author} {\bibinfo {author} {\bibfnamefont {L.~V.}\ \bibnamefont
  {Keldysh}},\ }\bibfield  {title} {\bibinfo {title} {Coulomb interaction in
  thin semiconductor and semimetal films},\ }\href
  {http://jetpletters.ru/ps/0/article_22207.shtml} {\bibfield  {journal}
  {\bibinfo  {journal} {JETP Lett.}\ }\textbf {\bibinfo {volume} {29}},\
  \bibinfo {pages} {658} (\bibinfo {year} {1979})}\BibitemShut {NoStop}%
\bibitem [{\citenamefont {Berkelbach}\ \emph {et~al.}(2013)\citenamefont
  {Berkelbach}, \citenamefont {Hybertsen},\ and\ \citenamefont
  {Reichman}}]{berkelbach2013theory}%
  \BibitemOpen
  \bibfield  {author} {\bibinfo {author} {\bibfnamefont {T.~C.}\ \bibnamefont
  {Berkelbach}}, \bibinfo {author} {\bibfnamefont {M.~S.}\ \bibnamefont
  {Hybertsen}},\ and\ \bibinfo {author} {\bibfnamefont {D.~R.}\ \bibnamefont
  {Reichman}},\ }\bibfield  {title} {\bibinfo {title} {Theory of neutral and
  charged excitons in monolayer transition metal dichalcogenides},\ }\href
  {https://journals.aps.org/prb/abstract/10.1103/PhysRevB.88.045318} {\bibfield
   {journal} {\bibinfo  {journal} {Physical Review B}\ }\textbf {\bibinfo
  {volume} {88}},\ \bibinfo {pages} {045318} (\bibinfo {year}
  {2013})}\BibitemShut {NoStop}%
\bibitem [{\citenamefont {H{\"u}ser}\ \emph {et~al.}(2013)\citenamefont
  {H{\"u}ser}, \citenamefont {Olsen},\ and\ \citenamefont
  {Thygesen}}]{huser2013dielectric}%
  \BibitemOpen
  \bibfield  {author} {\bibinfo {author} {\bibfnamefont {F.}~\bibnamefont
  {H{\"u}ser}}, \bibinfo {author} {\bibfnamefont {T.}~\bibnamefont {Olsen}},\
  and\ \bibinfo {author} {\bibfnamefont {K.~S.}\ \bibnamefont {Thygesen}},\
  }\bibfield  {title} {\bibinfo {title} {How dielectric screening in
  two-dimensional crystals affects the convergence of excited-state
  calculations: Monolayer mos 2},\ }\href
  {https://journals.aps.org/prb/abstract/10.1103/PhysRevB.88.245309} {\bibfield
   {journal} {\bibinfo  {journal} {Physical Review B}\ }\textbf {\bibinfo
  {volume} {88}},\ \bibinfo {pages} {245309} (\bibinfo {year}
  {2013})}\BibitemShut {NoStop}%
\bibitem [{\citenamefont {Korm{\'a}nyos}\ \emph {et~al.}(2015)\citenamefont
  {Korm{\'a}nyos}, \citenamefont {Burkard}, \citenamefont {Gmitra},
  \citenamefont {Fabian}, \citenamefont {Z{\'o}lyomi}, \citenamefont
  {Drummond},\ and\ \citenamefont {Fal’ko}}]{kormanyos2015k}%
  \BibitemOpen
  \bibfield  {author} {\bibinfo {author} {\bibfnamefont {A.}~\bibnamefont
  {Korm{\'a}nyos}}, \bibinfo {author} {\bibfnamefont {G.}~\bibnamefont
  {Burkard}}, \bibinfo {author} {\bibfnamefont {M.}~\bibnamefont {Gmitra}},
  \bibinfo {author} {\bibfnamefont {J.}~\bibnamefont {Fabian}}, \bibinfo
  {author} {\bibfnamefont {V.}~\bibnamefont {Z{\'o}lyomi}}, \bibinfo {author}
  {\bibfnamefont {N.~D.}\ \bibnamefont {Drummond}},\ and\ \bibinfo {author}
  {\bibfnamefont {V.}~\bibnamefont {Fal’ko}},\ }\bibfield  {title} {\bibinfo
  {title} {k{\textperiodcentered} p theory for two-dimensional transition metal
  dichalcogenide semiconductors},\ }\href
  {https://iopscience.iop.org/article/10.1088/2053-1583/2/2/022001} {\bibfield
  {journal} {\bibinfo  {journal} {2D Materials}\ }\textbf {\bibinfo {volume}
  {2}},\ \bibinfo {pages} {022001} (\bibinfo {year} {2015})}\BibitemShut
  {NoStop}%
\bibitem [{\citenamefont {Srivastava}\ \emph {et~al.}(2015)\citenamefont
  {Srivastava}, \citenamefont {Sidler}, \citenamefont {Allain}, \citenamefont
  {Lembke}, \citenamefont {Kis},\ and\ \citenamefont
  {Imamo{\u{g}}lu}}]{srivastava2015valley}%
  \BibitemOpen
  \bibfield  {author} {\bibinfo {author} {\bibfnamefont {A.}~\bibnamefont
  {Srivastava}}, \bibinfo {author} {\bibfnamefont {M.}~\bibnamefont {Sidler}},
  \bibinfo {author} {\bibfnamefont {A.~V.}\ \bibnamefont {Allain}}, \bibinfo
  {author} {\bibfnamefont {D.~S.}\ \bibnamefont {Lembke}}, \bibinfo {author}
  {\bibfnamefont {A.}~\bibnamefont {Kis}},\ and\ \bibinfo {author}
  {\bibfnamefont {A.}~\bibnamefont {Imamo{\u{g}}lu}},\ }\bibfield  {title}
  {\bibinfo {title} {Valley zeeman effect in elementary optical excitations of
  monolayer wse 2},\ }\href {https://www.nature.com/articles/nphys3203}
  {\bibfield  {journal} {\bibinfo  {journal} {Nature Physics}\ }\textbf
  {\bibinfo {volume} {11}},\ \bibinfo {pages} {141} (\bibinfo {year}
  {2015})}\BibitemShut {NoStop}%
\bibitem [{\citenamefont {Mostaani}\ \emph {et~al.}(2017)\citenamefont
  {Mostaani}, \citenamefont {Szyniszewski}, \citenamefont {Price},
  \citenamefont {Maezono}, \citenamefont {Danovich}, \citenamefont {Hunt},
  \citenamefont {Drummond},\ and\ \citenamefont
  {Fal'Ko}}]{mostaani2017diffusion}%
  \BibitemOpen
  \bibfield  {author} {\bibinfo {author} {\bibfnamefont {E.}~\bibnamefont
  {Mostaani}}, \bibinfo {author} {\bibfnamefont {M.}~\bibnamefont
  {Szyniszewski}}, \bibinfo {author} {\bibfnamefont {C.}~\bibnamefont {Price}},
  \bibinfo {author} {\bibfnamefont {R.}~\bibnamefont {Maezono}}, \bibinfo
  {author} {\bibfnamefont {M.}~\bibnamefont {Danovich}}, \bibinfo {author}
  {\bibfnamefont {R.}~\bibnamefont {Hunt}}, \bibinfo {author} {\bibfnamefont
  {N.}~\bibnamefont {Drummond}},\ and\ \bibinfo {author} {\bibfnamefont
  {V.}~\bibnamefont {Fal'Ko}},\ }\bibfield  {title} {\bibinfo {title}
  {Diffusion quantum monte carlo study of excitonic complexes in
  two-dimensional transition-metal dichalcogenides},\ }\href
  {https://journals.aps.org/prb/abstract/10.1103/PhysRevB.96.075431} {\bibfield
   {journal} {\bibinfo  {journal} {Physical Review B}\ }\textbf {\bibinfo
  {volume} {96}},\ \bibinfo {pages} {075431} (\bibinfo {year}
  {2017})}\BibitemShut {NoStop}%
\bibitem [{\citenamefont {Mueller}\ and\ \citenamefont
  {Malic}(2018)}]{mueller2018exciton}%
  \BibitemOpen
  \bibfield  {author} {\bibinfo {author} {\bibfnamefont {T.}~\bibnamefont
  {Mueller}}\ and\ \bibinfo {author} {\bibfnamefont {E.}~\bibnamefont
  {Malic}},\ }\bibfield  {title} {\bibinfo {title} {Exciton physics and device
  application of two-dimensional transition metal dichalcogenide
  semiconductors},\ }\href {https://www.nature.com/articles/s41699-018-0074-2}
  {\bibfield  {journal} {\bibinfo  {journal} {npj 2D Materials and
  Applications}\ }\textbf {\bibinfo {volume} {2}},\ \bibinfo {pages} {29}
  (\bibinfo {year} {2018})}\BibitemShut {NoStop}%
\bibitem [{\citenamefont {Anantharaman}\ \emph {et~al.}(2021)\citenamefont
  {Anantharaman}, \citenamefont {Jo},\ and\ \citenamefont
  {Jariwala}}]{anantharaman2021exciton}%
  \BibitemOpen
  \bibfield  {author} {\bibinfo {author} {\bibfnamefont {S.~B.}\ \bibnamefont
  {Anantharaman}}, \bibinfo {author} {\bibfnamefont {K.}~\bibnamefont {Jo}},\
  and\ \bibinfo {author} {\bibfnamefont {D.}~\bibnamefont {Jariwala}},\
  }\bibfield  {title} {\bibinfo {title} {Exciton--photonics: from fundamental
  science to applications},\ }\href
  {https://pubs.acs.org/doi/10.1021/acsnano.1c02204} {\bibfield  {journal}
  {\bibinfo  {journal} {ACS nano}\ }\textbf {\bibinfo {volume} {15}},\ \bibinfo
  {pages} {12628} (\bibinfo {year} {2021})}\BibitemShut {NoStop}%
\bibitem [{\citenamefont {Ciarrocchi}\ \emph {et~al.}(2022)\citenamefont
  {Ciarrocchi}, \citenamefont {Tagarelli}, \citenamefont {Avsar},\ and\
  \citenamefont {Kis}}]{ciarrocchi2022excitonic}%
  \BibitemOpen
  \bibfield  {author} {\bibinfo {author} {\bibfnamefont {A.}~\bibnamefont
  {Ciarrocchi}}, \bibinfo {author} {\bibfnamefont {F.}~\bibnamefont
  {Tagarelli}}, \bibinfo {author} {\bibfnamefont {A.}~\bibnamefont {Avsar}},\
  and\ \bibinfo {author} {\bibfnamefont {A.}~\bibnamefont {Kis}},\ }\bibfield
  {title} {\bibinfo {title} {Excitonic devices with van der waals
  heterostructures: valleytronics meets twistronics},\ }\href
  {https://www.nature.com/articles/s41578-021-00408-7} {\bibfield  {journal}
  {\bibinfo  {journal} {Nature Reviews Materials}\ }\textbf {\bibinfo {volume}
  {7}},\ \bibinfo {pages} {449} (\bibinfo {year} {2022})}\BibitemShut {NoStop}%
\bibitem [{\citenamefont {Chernikov}\ \emph
  {et~al.}(2015{\natexlab{a}})\citenamefont {Chernikov}, \citenamefont
  {Ruppert}, \citenamefont {Hill}, \citenamefont {Rigosi},\ and\ \citenamefont
  {Heinz}}]{chernikov2015population}%
  \BibitemOpen
  \bibfield  {author} {\bibinfo {author} {\bibfnamefont {A.}~\bibnamefont
  {Chernikov}}, \bibinfo {author} {\bibfnamefont {C.}~\bibnamefont {Ruppert}},
  \bibinfo {author} {\bibfnamefont {H.~M.}\ \bibnamefont {Hill}}, \bibinfo
  {author} {\bibfnamefont {A.~F.}\ \bibnamefont {Rigosi}},\ and\ \bibinfo
  {author} {\bibfnamefont {T.~F.}\ \bibnamefont {Heinz}},\ }\bibfield  {title}
  {\bibinfo {title} {Population inversion and giant bandgap renormalization in
  atomically thin ws2 layers},\ }\href
  {https://www.nature.com/articles/nphoton.2015.104} {\bibfield  {journal}
  {\bibinfo  {journal} {Nature Photonics}\ }\textbf {\bibinfo {volume} {9}},\
  \bibinfo {pages} {466} (\bibinfo {year} {2015}{\natexlab{a}})}\BibitemShut
  {NoStop}%
\bibitem [{\citenamefont {Steinhoff}\ \emph {et~al.}(2017)\citenamefont
  {Steinhoff}, \citenamefont {Florian}, \citenamefont {R{\"o}sner},
  \citenamefont {Sch{\"o}nhoff}, \citenamefont {Wehling},\ and\ \citenamefont
  {Jahnke}}]{steinhoff2017exciton}%
  \BibitemOpen
  \bibfield  {author} {\bibinfo {author} {\bibfnamefont {A.}~\bibnamefont
  {Steinhoff}}, \bibinfo {author} {\bibfnamefont {M.}~\bibnamefont {Florian}},
  \bibinfo {author} {\bibfnamefont {M.}~\bibnamefont {R{\"o}sner}}, \bibinfo
  {author} {\bibfnamefont {G.}~\bibnamefont {Sch{\"o}nhoff}}, \bibinfo {author}
  {\bibfnamefont {T.~O.}\ \bibnamefont {Wehling}},\ and\ \bibinfo {author}
  {\bibfnamefont {F.}~\bibnamefont {Jahnke}},\ }\bibfield  {title} {\bibinfo
  {title} {Exciton fission in monolayer transition metal dichalcogenide
  semiconductors},\ }\href {https://www.nature.com/articles/s41467-017-01298-6}
  {\bibfield  {journal} {\bibinfo  {journal} {Nature communications}\ }\textbf
  {\bibinfo {volume} {8}},\ \bibinfo {pages} {1166} (\bibinfo {year}
  {2017})}\BibitemShut {NoStop}%
\bibitem [{\citenamefont {Estrecho}\ \emph {et~al.}(2019)\citenamefont
  {Estrecho}, \citenamefont {Gao}, \citenamefont {Bobrovska}, \citenamefont
  {Comber-Todd}, \citenamefont {Fraser}, \citenamefont {Steger}, \citenamefont
  {West}, \citenamefont {Pfeiffer}, \citenamefont {Levinsen}, \citenamefont
  {Parish} \emph {et~al.}}]{estrecho2019direct}%
  \BibitemOpen
  \bibfield  {author} {\bibinfo {author} {\bibfnamefont {E.}~\bibnamefont
  {Estrecho}}, \bibinfo {author} {\bibfnamefont {T.}~\bibnamefont {Gao}},
  \bibinfo {author} {\bibfnamefont {N.}~\bibnamefont {Bobrovska}}, \bibinfo
  {author} {\bibfnamefont {D.}~\bibnamefont {Comber-Todd}}, \bibinfo {author}
  {\bibfnamefont {M.~D.}\ \bibnamefont {Fraser}}, \bibinfo {author}
  {\bibfnamefont {M.}~\bibnamefont {Steger}}, \bibinfo {author} {\bibfnamefont
  {K.}~\bibnamefont {West}}, \bibinfo {author} {\bibfnamefont {L.~N.}\
  \bibnamefont {Pfeiffer}}, \bibinfo {author} {\bibfnamefont {J.}~\bibnamefont
  {Levinsen}}, \bibinfo {author} {\bibfnamefont {M.}~\bibnamefont {Parish}},
  \emph {et~al.},\ }\bibfield  {title} {\bibinfo {title} {Direct measurement of
  polariton-polariton interaction strength in the thomas-fermi regime of
  exciton-polariton condensation},\ }\href@noop {} {\bibfield  {journal}
  {\bibinfo  {journal} {Physical Review B}\ }\textbf {\bibinfo {volume}
  {100}},\ \bibinfo {pages} {035306} (\bibinfo {year} {2019})}\BibitemShut
  {NoStop}%
\bibitem [{\citenamefont {Shahnazaryan}\ \emph {et~al.}(2017)\citenamefont
  {Shahnazaryan}, \citenamefont {Iorsh}, \citenamefont {Shelykh},\ and\
  \citenamefont {Kyriienko}}]{shahnazaryan2017exciton}%
  \BibitemOpen
  \bibfield  {author} {\bibinfo {author} {\bibfnamefont {V.}~\bibnamefont
  {Shahnazaryan}}, \bibinfo {author} {\bibfnamefont {I.}~\bibnamefont {Iorsh}},
  \bibinfo {author} {\bibfnamefont {I.~A.}\ \bibnamefont {Shelykh}},\ and\
  \bibinfo {author} {\bibfnamefont {O.}~\bibnamefont {Kyriienko}},\ }\bibfield
  {title} {\bibinfo {title} {Exciton-exciton interaction in transition-metal
  dichalcogenide monolayers},\ }\href
  {https://journals.aps.org/prb/abstract/10.1103/PhysRevB.96.115409} {\bibfield
   {journal} {\bibinfo  {journal} {Physical Review B}\ }\textbf {\bibinfo
  {volume} {96}},\ \bibinfo {pages} {115409} (\bibinfo {year}
  {2017})}\BibitemShut {NoStop}%
\bibitem [{\citenamefont {Shahnazaryan}\ \emph {et~al.}(2020)\citenamefont
  {Shahnazaryan}, \citenamefont {Kozin}, \citenamefont {Shelykh}, \citenamefont
  {Iorsh},\ and\ \citenamefont {Kyriienko}}]{shahnazaryan2020tunable}%
  \BibitemOpen
  \bibfield  {author} {\bibinfo {author} {\bibfnamefont {V.}~\bibnamefont
  {Shahnazaryan}}, \bibinfo {author} {\bibfnamefont {V.}~\bibnamefont {Kozin}},
  \bibinfo {author} {\bibfnamefont {I.}~\bibnamefont {Shelykh}}, \bibinfo
  {author} {\bibfnamefont {I.}~\bibnamefont {Iorsh}},\ and\ \bibinfo {author}
  {\bibfnamefont {O.}~\bibnamefont {Kyriienko}},\ }\bibfield  {title} {\bibinfo
  {title} {Tunable optical nonlinearity for transition metal dichalcogenide
  polaritons dressed by a fermi sea},\ }\href
  {https://journals.aps.org/prb/abstract/10.1103/PhysRevB.102.115310}
  {\bibfield  {journal} {\bibinfo  {journal} {Physical Review B}\ }\textbf
  {\bibinfo {volume} {102}},\ \bibinfo {pages} {115310} (\bibinfo {year}
  {2020})}\BibitemShut {NoStop}%
\bibitem [{\citenamefont {Bleu}\ \emph {et~al.}(2020)\citenamefont {Bleu},
  \citenamefont {Li}, \citenamefont {Levinsen},\ and\ \citenamefont
  {Parish}}]{bleu2020polariton}%
  \BibitemOpen
  \bibfield  {author} {\bibinfo {author} {\bibfnamefont {O.}~\bibnamefont
  {Bleu}}, \bibinfo {author} {\bibfnamefont {G.}~\bibnamefont {Li}}, \bibinfo
  {author} {\bibfnamefont {J.}~\bibnamefont {Levinsen}},\ and\ \bibinfo
  {author} {\bibfnamefont {M.~M.}\ \bibnamefont {Parish}},\ }\bibfield  {title}
  {\bibinfo {title} {Polariton interactions in microcavities with atomically
  thin semiconductor layers},\ }\href
  {https://journals.aps.org/prresearch/abstract/10.1103/PhysRevResearch.2.043185}
  {\bibfield  {journal} {\bibinfo  {journal} {Physical Review Research}\
  }\textbf {\bibinfo {volume} {2}},\ \bibinfo {pages} {043185} (\bibinfo {year}
  {2020})}\BibitemShut {NoStop}%
\bibitem [{\citenamefont {Fey}\ \emph {et~al.}(2020)\citenamefont {Fey},
  \citenamefont {Schmelcher}, \citenamefont {Imamoglu},\ and\ \citenamefont
  {Schmidt}}]{fey2020theory}%
  \BibitemOpen
  \bibfield  {author} {\bibinfo {author} {\bibfnamefont {C.}~\bibnamefont
  {Fey}}, \bibinfo {author} {\bibfnamefont {P.}~\bibnamefont {Schmelcher}},
  \bibinfo {author} {\bibfnamefont {A.}~\bibnamefont {Imamoglu}},\ and\
  \bibinfo {author} {\bibfnamefont {R.}~\bibnamefont {Schmidt}},\ }\bibfield
  {title} {\bibinfo {title} {Theory of exciton-electron scattering in
  atomically thin semiconductors},\ }\href
  {https://journals.aps.org/prb/abstract/10.1103/PhysRevB.101.195417}
  {\bibfield  {journal} {\bibinfo  {journal} {Physical Review B}\ }\textbf
  {\bibinfo {volume} {101}},\ \bibinfo {pages} {195417} (\bibinfo {year}
  {2020})}\BibitemShut {NoStop}%
\bibitem [{\citenamefont {Erkensten}\ \emph {et~al.}(2021)\citenamefont
  {Erkensten}, \citenamefont {Brem},\ and\ \citenamefont
  {Malic}}]{erkensten2021exciton}%
  \BibitemOpen
  \bibfield  {author} {\bibinfo {author} {\bibfnamefont {D.}~\bibnamefont
  {Erkensten}}, \bibinfo {author} {\bibfnamefont {S.}~\bibnamefont {Brem}},\
  and\ \bibinfo {author} {\bibfnamefont {E.}~\bibnamefont {Malic}},\ }\bibfield
   {title} {\bibinfo {title} {Exciton-exciton interaction in transition metal
  dichalcogenide monolayers and van der waals heterostructures},\ }\href
  {https://journals.aps.org/prb/abstract/10.1103/PhysRevB.103.045426}
  {\bibfield  {journal} {\bibinfo  {journal} {Physical Review B}\ }\textbf
  {\bibinfo {volume} {103}},\ \bibinfo {pages} {045426} (\bibinfo {year}
  {2021})}\BibitemShut {NoStop}%
\bibitem [{\citenamefont {Cam}\ \emph {et~al.}(2022)\citenamefont {Cam},
  \citenamefont {Phuc},\ and\ \citenamefont {Osipov}}]{cam2022symmetry}%
  \BibitemOpen
  \bibfield  {author} {\bibinfo {author} {\bibfnamefont {H.~N.}\ \bibnamefont
  {Cam}}, \bibinfo {author} {\bibfnamefont {N.~T.}\ \bibnamefont {Phuc}},\ and\
  \bibinfo {author} {\bibfnamefont {V.~A.}\ \bibnamefont {Osipov}},\ }\bibfield
   {title} {\bibinfo {title} {Symmetry-dependent exciton-exciton interaction
  and intervalley biexciton in monolayer transition metal dichalcogenides},\
  }\href {https://www.nature.com/articles/s41699-022-00290-z} {\bibfield
  {journal} {\bibinfo  {journal} {npj 2D Materials and Applications}\ }\textbf
  {\bibinfo {volume} {6}},\ \bibinfo {pages} {22} (\bibinfo {year}
  {2022})}\BibitemShut {NoStop}%
\bibitem [{\citenamefont {Dufferwiel}\ \emph {et~al.}(2015)\citenamefont
  {Dufferwiel}, \citenamefont {Schwarz}, \citenamefont {Withers}, \citenamefont
  {Trichet}, \citenamefont {Li}, \citenamefont {Sich}, \citenamefont {Del
  Pozo-Zamudio}, \citenamefont {Clark}, \citenamefont {Nalitov}, \citenamefont
  {Solnyshkov} \emph {et~al.}}]{dufferwiel2015exciton}%
  \BibitemOpen
  \bibfield  {author} {\bibinfo {author} {\bibfnamefont {S.}~\bibnamefont
  {Dufferwiel}}, \bibinfo {author} {\bibfnamefont {S.}~\bibnamefont {Schwarz}},
  \bibinfo {author} {\bibfnamefont {F.}~\bibnamefont {Withers}}, \bibinfo
  {author} {\bibfnamefont {A.}~\bibnamefont {Trichet}}, \bibinfo {author}
  {\bibfnamefont {F.}~\bibnamefont {Li}}, \bibinfo {author} {\bibfnamefont
  {M.}~\bibnamefont {Sich}}, \bibinfo {author} {\bibfnamefont {O.}~\bibnamefont
  {Del Pozo-Zamudio}}, \bibinfo {author} {\bibfnamefont {C.}~\bibnamefont
  {Clark}}, \bibinfo {author} {\bibfnamefont {A.}~\bibnamefont {Nalitov}},
  \bibinfo {author} {\bibfnamefont {D.}~\bibnamefont {Solnyshkov}}, \emph
  {et~al.},\ }\bibfield  {title} {\bibinfo {title} {Exciton--polaritons in van
  der waals heterostructures embedded in tunable microcavities},\ }\href
  {https://www.nature.com/articles/ncomms9579} {\bibfield  {journal} {\bibinfo
  {journal} {Nature communications}\ }\textbf {\bibinfo {volume} {6}},\
  \bibinfo {pages} {8579} (\bibinfo {year} {2015})}\BibitemShut {NoStop}%
\bibitem [{\citenamefont {Chernikov}\ \emph
  {et~al.}(2015{\natexlab{b}})\citenamefont {Chernikov}, \citenamefont {Van
  Der~Zande}, \citenamefont {Hill}, \citenamefont {Rigosi}, \citenamefont
  {Velauthapillai}, \citenamefont {Hone},\ and\ \citenamefont
  {Heinz}}]{chernikov2015electrical}%
  \BibitemOpen
  \bibfield  {author} {\bibinfo {author} {\bibfnamefont {A.}~\bibnamefont
  {Chernikov}}, \bibinfo {author} {\bibfnamefont {A.~M.}\ \bibnamefont {Van
  Der~Zande}}, \bibinfo {author} {\bibfnamefont {H.~M.}\ \bibnamefont {Hill}},
  \bibinfo {author} {\bibfnamefont {A.~F.}\ \bibnamefont {Rigosi}}, \bibinfo
  {author} {\bibfnamefont {A.}~\bibnamefont {Velauthapillai}}, \bibinfo
  {author} {\bibfnamefont {J.}~\bibnamefont {Hone}},\ and\ \bibinfo {author}
  {\bibfnamefont {T.~F.}\ \bibnamefont {Heinz}},\ }\bibfield  {title} {\bibinfo
  {title} {Electrical tuning of exciton binding energies in monolayer ws 2},\
  }\href {https://journals.aps.org/prl/abstract/10.1103/PhysRevLett.115.126802}
  {\bibfield  {journal} {\bibinfo  {journal} {Physical review letters}\
  }\textbf {\bibinfo {volume} {115}},\ \bibinfo {pages} {126802} (\bibinfo
  {year} {2015}{\natexlab{b}})}\BibitemShut {NoStop}%
\bibitem [{\citenamefont {Mak}\ \emph {et~al.}(2013)\citenamefont {Mak},
  \citenamefont {He}, \citenamefont {Lee}, \citenamefont {Lee}, \citenamefont
  {Hone}, \citenamefont {Heinz},\ and\ \citenamefont {Shan}}]{mak2013tightly}%
  \BibitemOpen
  \bibfield  {author} {\bibinfo {author} {\bibfnamefont {K.~F.}\ \bibnamefont
  {Mak}}, \bibinfo {author} {\bibfnamefont {K.}~\bibnamefont {He}}, \bibinfo
  {author} {\bibfnamefont {C.}~\bibnamefont {Lee}}, \bibinfo {author}
  {\bibfnamefont {G.~H.}\ \bibnamefont {Lee}}, \bibinfo {author} {\bibfnamefont
  {J.}~\bibnamefont {Hone}}, \bibinfo {author} {\bibfnamefont {T.~F.}\
  \bibnamefont {Heinz}},\ and\ \bibinfo {author} {\bibfnamefont
  {J.}~\bibnamefont {Shan}},\ }\bibfield  {title} {\bibinfo {title} {Tightly
  bound trions in monolayer mos2},\ }\href
  {https://www.nature.com/articles/nmat3505} {\bibfield  {journal} {\bibinfo
  {journal} {Nature materials}\ }\textbf {\bibinfo {volume} {12}},\ \bibinfo
  {pages} {207} (\bibinfo {year} {2013})}\BibitemShut {NoStop}%
\bibitem [{\citenamefont {Ross}\ \emph {et~al.}(2013)\citenamefont {Ross},
  \citenamefont {Wu}, \citenamefont {Yu}, \citenamefont {Ghimire},
  \citenamefont {Jones}, \citenamefont {Aivazian}, \citenamefont {Yan},
  \citenamefont {Mandrus}, \citenamefont {Xiao}, \citenamefont {Yao} \emph
  {et~al.}}]{ross2013electrical}%
  \BibitemOpen
  \bibfield  {author} {\bibinfo {author} {\bibfnamefont {J.~S.}\ \bibnamefont
  {Ross}}, \bibinfo {author} {\bibfnamefont {S.}~\bibnamefont {Wu}}, \bibinfo
  {author} {\bibfnamefont {H.}~\bibnamefont {Yu}}, \bibinfo {author}
  {\bibfnamefont {N.~J.}\ \bibnamefont {Ghimire}}, \bibinfo {author}
  {\bibfnamefont {A.~M.}\ \bibnamefont {Jones}}, \bibinfo {author}
  {\bibfnamefont {G.}~\bibnamefont {Aivazian}}, \bibinfo {author}
  {\bibfnamefont {J.}~\bibnamefont {Yan}}, \bibinfo {author} {\bibfnamefont
  {D.~G.}\ \bibnamefont {Mandrus}}, \bibinfo {author} {\bibfnamefont
  {D.}~\bibnamefont {Xiao}}, \bibinfo {author} {\bibfnamefont {W.}~\bibnamefont
  {Yao}}, \emph {et~al.},\ }\bibfield  {title} {\bibinfo {title} {Electrical
  control of neutral and charged excitons in a monolayer semiconductor},\
  }\href {https://www.nature.com/articles/ncomms2498} {\bibfield  {journal}
  {\bibinfo  {journal} {Nature communications}\ }\textbf {\bibinfo {volume}
  {4}},\ \bibinfo {pages} {1474} (\bibinfo {year} {2013})}\BibitemShut
  {NoStop}%
\bibitem [{\citenamefont {Singh}\ \emph {et~al.}(2016)\citenamefont {Singh},
  \citenamefont {Moody}, \citenamefont {Tran}, \citenamefont {Scott},
  \citenamefont {Overbeck}, \citenamefont {Bergh{\"a}user}, \citenamefont
  {Schaibley}, \citenamefont {Seifert}, \citenamefont {Pleskot}, \citenamefont
  {Gabor} \emph {et~al.}}]{singh2016trion}%
  \BibitemOpen
  \bibfield  {author} {\bibinfo {author} {\bibfnamefont {A.}~\bibnamefont
  {Singh}}, \bibinfo {author} {\bibfnamefont {G.}~\bibnamefont {Moody}},
  \bibinfo {author} {\bibfnamefont {K.}~\bibnamefont {Tran}}, \bibinfo {author}
  {\bibfnamefont {M.~E.}\ \bibnamefont {Scott}}, \bibinfo {author}
  {\bibfnamefont {V.}~\bibnamefont {Overbeck}}, \bibinfo {author}
  {\bibfnamefont {G.}~\bibnamefont {Bergh{\"a}user}}, \bibinfo {author}
  {\bibfnamefont {J.}~\bibnamefont {Schaibley}}, \bibinfo {author}
  {\bibfnamefont {E.~J.}\ \bibnamefont {Seifert}}, \bibinfo {author}
  {\bibfnamefont {D.}~\bibnamefont {Pleskot}}, \bibinfo {author} {\bibfnamefont
  {N.~M.}\ \bibnamefont {Gabor}}, \emph {et~al.},\ }\bibfield  {title}
  {\bibinfo {title} {Trion formation dynamics in monolayer transition metal
  dichalcogenides},\ }\href
  {https://journals.aps.org/prb/abstract/10.1103/PhysRevB.93.041401} {\bibfield
   {journal} {\bibinfo  {journal} {Physical Review B}\ }\textbf {\bibinfo
  {volume} {93}},\ \bibinfo {pages} {041401} (\bibinfo {year}
  {2016})}\BibitemShut {NoStop}%
\bibitem [{\citenamefont {Courtade}\ \emph {et~al.}(2017)\citenamefont
  {Courtade}, \citenamefont {Semina}, \citenamefont {Manca}, \citenamefont
  {Glazov}, \citenamefont {Robert}, \citenamefont {Cadiz}, \citenamefont
  {Wang}, \citenamefont {Taniguchi}, \citenamefont {Watanabe}, \citenamefont
  {Pierre} \emph {et~al.}}]{courtade2017charged}%
  \BibitemOpen
  \bibfield  {author} {\bibinfo {author} {\bibfnamefont {E.}~\bibnamefont
  {Courtade}}, \bibinfo {author} {\bibfnamefont {M.}~\bibnamefont {Semina}},
  \bibinfo {author} {\bibfnamefont {M.}~\bibnamefont {Manca}}, \bibinfo
  {author} {\bibfnamefont {M.}~\bibnamefont {Glazov}}, \bibinfo {author}
  {\bibfnamefont {C.}~\bibnamefont {Robert}}, \bibinfo {author} {\bibfnamefont
  {F.}~\bibnamefont {Cadiz}}, \bibinfo {author} {\bibfnamefont
  {G.}~\bibnamefont {Wang}}, \bibinfo {author} {\bibfnamefont {T.}~\bibnamefont
  {Taniguchi}}, \bibinfo {author} {\bibfnamefont {K.}~\bibnamefont {Watanabe}},
  \bibinfo {author} {\bibfnamefont {M.}~\bibnamefont {Pierre}}, \emph
  {et~al.},\ }\bibfield  {title} {\bibinfo {title} {Charged excitons in
  monolayer wse 2: Experiment and theory},\ }\href
  {https://journals.aps.org/prb/abstract/10.1103/PhysRevB.96.085302} {\bibfield
   {journal} {\bibinfo  {journal} {Physical Review B}\ }\textbf {\bibinfo
  {volume} {96}},\ \bibinfo {pages} {085302} (\bibinfo {year}
  {2017})}\BibitemShut {NoStop}%
\bibitem [{\citenamefont {Sidler}\ \emph {et~al.}(2017)\citenamefont {Sidler},
  \citenamefont {Back}, \citenamefont {Cotlet}, \citenamefont {Srivastava},
  \citenamefont {Fink}, \citenamefont {Kroner}, \citenamefont {Demler},\ and\
  \citenamefont {Imamoglu}}]{sidler2017fermi}%
  \BibitemOpen
  \bibfield  {author} {\bibinfo {author} {\bibfnamefont {M.}~\bibnamefont
  {Sidler}}, \bibinfo {author} {\bibfnamefont {P.}~\bibnamefont {Back}},
  \bibinfo {author} {\bibfnamefont {O.}~\bibnamefont {Cotlet}}, \bibinfo
  {author} {\bibfnamefont {A.}~\bibnamefont {Srivastava}}, \bibinfo {author}
  {\bibfnamefont {T.}~\bibnamefont {Fink}}, \bibinfo {author} {\bibfnamefont
  {M.}~\bibnamefont {Kroner}}, \bibinfo {author} {\bibfnamefont
  {E.}~\bibnamefont {Demler}},\ and\ \bibinfo {author} {\bibfnamefont
  {A.}~\bibnamefont {Imamoglu}},\ }\bibfield  {title} {\bibinfo {title} {Fermi
  polaron-polaritons in charge-tunable atomically thin semiconductors},\ }\href
  {https://www.nature.com/articles/nphys3949} {\bibfield  {journal} {\bibinfo
  {journal} {Nature Physics}\ }\textbf {\bibinfo {volume} {13}},\ \bibinfo
  {pages} {255} (\bibinfo {year} {2017})}\BibitemShut {NoStop}%
\bibitem [{\citenamefont {Efimkin}\ and\ \citenamefont
  {MacDonald}(2017)}]{efimkin2017many}%
  \BibitemOpen
  \bibfield  {author} {\bibinfo {author} {\bibfnamefont {D.~K.}\ \bibnamefont
  {Efimkin}}\ and\ \bibinfo {author} {\bibfnamefont {A.~H.}\ \bibnamefont
  {MacDonald}},\ }\bibfield  {title} {\bibinfo {title} {Many-body theory of
  trion absorption features in two-dimensional semiconductors},\ }\href
  {https://journals.aps.org/prb/abstract/10.1103/PhysRevB.95.035417} {\bibfield
   {journal} {\bibinfo  {journal} {Physical Review B}\ }\textbf {\bibinfo
  {volume} {95}},\ \bibinfo {pages} {035417} (\bibinfo {year}
  {2017})}\BibitemShut {NoStop}%
\bibitem [{\citenamefont {Tan}\ \emph {et~al.}(2020)\citenamefont {Tan},
  \citenamefont {Cotlet}, \citenamefont {Bergschneider}, \citenamefont
  {Schmidt}, \citenamefont {Back}, \citenamefont {Shimazaki}, \citenamefont
  {Kroner},\ and\ \citenamefont {{\.I}mamo{\u{g}}lu}}]{tan2020interacting}%
  \BibitemOpen
  \bibfield  {author} {\bibinfo {author} {\bibfnamefont {L.~B.}\ \bibnamefont
  {Tan}}, \bibinfo {author} {\bibfnamefont {O.}~\bibnamefont {Cotlet}},
  \bibinfo {author} {\bibfnamefont {A.}~\bibnamefont {Bergschneider}}, \bibinfo
  {author} {\bibfnamefont {R.}~\bibnamefont {Schmidt}}, \bibinfo {author}
  {\bibfnamefont {P.}~\bibnamefont {Back}}, \bibinfo {author} {\bibfnamefont
  {Y.}~\bibnamefont {Shimazaki}}, \bibinfo {author} {\bibfnamefont
  {M.}~\bibnamefont {Kroner}},\ and\ \bibinfo {author} {\bibfnamefont
  {A.}~\bibnamefont {{\.I}mamo{\u{g}}lu}},\ }\bibfield  {title} {\bibinfo
  {title} {Interacting polaron-polaritons},\ }\href
  {https://journals.aps.org/prx/abstract/10.1103/PhysRevX.10.021011} {\bibfield
   {journal} {\bibinfo  {journal} {Physical Review X}\ }\textbf {\bibinfo
  {volume} {10}},\ \bibinfo {pages} {021011} (\bibinfo {year}
  {2020})}\BibitemShut {NoStop}%
\bibitem [{\citenamefont {Efimkin}\ \emph {et~al.}(2021)\citenamefont
  {Efimkin}, \citenamefont {Laird}, \citenamefont {Levinsen}, \citenamefont
  {Parish},\ and\ \citenamefont {MacDonald}}]{efimkin2021electron}%
  \BibitemOpen
  \bibfield  {author} {\bibinfo {author} {\bibfnamefont {D.~K.}\ \bibnamefont
  {Efimkin}}, \bibinfo {author} {\bibfnamefont {E.~K.}\ \bibnamefont {Laird}},
  \bibinfo {author} {\bibfnamefont {J.}~\bibnamefont {Levinsen}}, \bibinfo
  {author} {\bibfnamefont {M.~M.}\ \bibnamefont {Parish}},\ and\ \bibinfo
  {author} {\bibfnamefont {A.~H.}\ \bibnamefont {MacDonald}},\ }\bibfield
  {title} {\bibinfo {title} {Electron-exciton interactions in the
  exciton-polaron problem},\ }\href
  {https://journals.aps.org/prb/abstract/10.1103/PhysRevB.95.035417} {\bibfield
   {journal} {\bibinfo  {journal} {Physical Review B}\ }\textbf {\bibinfo
  {volume} {103}},\ \bibinfo {pages} {075417} (\bibinfo {year}
  {2021})}\BibitemShut {NoStop}%
\bibitem [{\citenamefont {Glazov}(2020)}]{glazov2020optical}%
  \BibitemOpen
  \bibfield  {author} {\bibinfo {author} {\bibfnamefont {M.~M.}\ \bibnamefont
  {Glazov}},\ }\bibfield  {title} {\bibinfo {title} {Optical properties of
  charged excitons in two-dimensional semiconductors},\ }\href
  {https://pubs.aip.org/aip/jcp/article-abstract/153/3/034703/1062654/Optical-properties-of-charged-excitons-in-two}
  {\bibfield  {journal} {\bibinfo  {journal} {The Journal of Chemical Physics}\
  }\textbf {\bibinfo {volume} {153}} (\bibinfo {year} {2020})}\BibitemShut
  {NoStop}%
\bibitem [{\citenamefont {Dufferwiel}\ \emph {et~al.}(2017)\citenamefont
  {Dufferwiel}, \citenamefont {Lyons}, \citenamefont {Solnyshkov},
  \citenamefont {Trichet}, \citenamefont {Withers}, \citenamefont {Schwarz},
  \citenamefont {Malpuech}, \citenamefont {Smith}, \citenamefont {Novoselov},
  \citenamefont {Skolnick} \emph {et~al.}}]{dufferwiel2017valley}%
  \BibitemOpen
  \bibfield  {author} {\bibinfo {author} {\bibfnamefont {S.}~\bibnamefont
  {Dufferwiel}}, \bibinfo {author} {\bibfnamefont {T.~P.}\ \bibnamefont
  {Lyons}}, \bibinfo {author} {\bibfnamefont {D.~D.}\ \bibnamefont
  {Solnyshkov}}, \bibinfo {author} {\bibfnamefont {A.~A.}\ \bibnamefont
  {Trichet}}, \bibinfo {author} {\bibfnamefont {F.}~\bibnamefont {Withers}},
  \bibinfo {author} {\bibfnamefont {S.}~\bibnamefont {Schwarz}}, \bibinfo
  {author} {\bibfnamefont {G.}~\bibnamefont {Malpuech}}, \bibinfo {author}
  {\bibfnamefont {J.~M.}\ \bibnamefont {Smith}}, \bibinfo {author}
  {\bibfnamefont {K.~S.}\ \bibnamefont {Novoselov}}, \bibinfo {author}
  {\bibfnamefont {M.~S.}\ \bibnamefont {Skolnick}}, \emph {et~al.},\ }\bibfield
   {title} {\bibinfo {title} {Valley-addressable polaritons in atomically thin
  semiconductors},\ }\href@noop {} {\bibfield  {journal} {\bibinfo  {journal}
  {Nature Photonics}\ }\textbf {\bibinfo {volume} {11}},\ \bibinfo {pages}
  {497} (\bibinfo {year} {2017})}\BibitemShut {NoStop}%
\bibitem [{\citenamefont {Zhumagulov}\ \emph {et~al.}(2022)\citenamefont
  {Zhumagulov}, \citenamefont {Chiavazzo}, \citenamefont {Gulevich},
  \citenamefont {Perebeinos}, \citenamefont {Shelykh},\ and\ \citenamefont
  {Kyriienko}}]{zhumagulov2022microscopic}%
  \BibitemOpen
  \bibfield  {author} {\bibinfo {author} {\bibfnamefont {Y.~V.}\ \bibnamefont
  {Zhumagulov}}, \bibinfo {author} {\bibfnamefont {S.}~\bibnamefont
  {Chiavazzo}}, \bibinfo {author} {\bibfnamefont {D.~R.}\ \bibnamefont
  {Gulevich}}, \bibinfo {author} {\bibfnamefont {V.}~\bibnamefont
  {Perebeinos}}, \bibinfo {author} {\bibfnamefont {I.~A.}\ \bibnamefont
  {Shelykh}},\ and\ \bibinfo {author} {\bibfnamefont {O.}~\bibnamefont
  {Kyriienko}},\ }\bibfield  {title} {\bibinfo {title} {Microscopic theory of
  exciton and trion polaritons in doped monolayers of transition metal
  dichalcogenides},\ }\href
  {https://www.nature.com/articles/s41524-022-00775-x} {\bibfield  {journal}
  {\bibinfo  {journal} {npj Computational Materials}\ }\textbf {\bibinfo
  {volume} {8}},\ \bibinfo {pages} {92} (\bibinfo {year} {2022})}\BibitemShut
  {NoStop}%
\bibitem [{\citenamefont {Emmanuele}\ \emph {et~al.}(2020)\citenamefont
  {Emmanuele}, \citenamefont {Sich}, \citenamefont {Kyriienko}, \citenamefont
  {Shahnazaryan}, \citenamefont {Withers}, \citenamefont {Catanzaro},
  \citenamefont {Walker}, \citenamefont {Benimetskiy}, \citenamefont
  {Skolnick}, \citenamefont {Tartakovskii} \emph
  {et~al.}}]{emmanuele2020highly}%
  \BibitemOpen
  \bibfield  {author} {\bibinfo {author} {\bibfnamefont {R.}~\bibnamefont
  {Emmanuele}}, \bibinfo {author} {\bibfnamefont {M.}~\bibnamefont {Sich}},
  \bibinfo {author} {\bibfnamefont {O.}~\bibnamefont {Kyriienko}}, \bibinfo
  {author} {\bibfnamefont {V.}~\bibnamefont {Shahnazaryan}}, \bibinfo {author}
  {\bibfnamefont {F.}~\bibnamefont {Withers}}, \bibinfo {author} {\bibfnamefont
  {A.}~\bibnamefont {Catanzaro}}, \bibinfo {author} {\bibfnamefont
  {P.}~\bibnamefont {Walker}}, \bibinfo {author} {\bibfnamefont
  {F.}~\bibnamefont {Benimetskiy}}, \bibinfo {author} {\bibfnamefont
  {M.}~\bibnamefont {Skolnick}}, \bibinfo {author} {\bibfnamefont
  {A.}~\bibnamefont {Tartakovskii}}, \emph {et~al.},\ }\bibfield  {title}
  {\bibinfo {title} {Highly nonlinear trion-polaritons in a monolayer
  semiconductor},\ }\href {https://www.nature.com/articles/s41467-020-17340-z}
  {\bibfield  {journal} {\bibinfo  {journal} {Nature communications}\ }\textbf
  {\bibinfo {volume} {11}},\ \bibinfo {pages} {3589} (\bibinfo {year}
  {2020})}\BibitemShut {NoStop}%
\bibitem [{\citenamefont {Kyriienko}\ \emph {et~al.}(2020)\citenamefont
  {Kyriienko}, \citenamefont {Krizhanovskii},\ and\ \citenamefont
  {Shelykh}}]{kyriienko2020nonlinear}%
  \BibitemOpen
  \bibfield  {author} {\bibinfo {author} {\bibfnamefont {O.}~\bibnamefont
  {Kyriienko}}, \bibinfo {author} {\bibfnamefont {D.}~\bibnamefont
  {Krizhanovskii}},\ and\ \bibinfo {author} {\bibfnamefont {I.}~\bibnamefont
  {Shelykh}},\ }\bibfield  {title} {\bibinfo {title} {Nonlinear quantum optics
  with trion polaritons in 2d monolayers: conventional and unconventional
  photon blockade},\ }\href
  {https://journals.aps.org/prl/abstract/10.1103/PhysRevLett.125.197402}
  {\bibfield  {journal} {\bibinfo  {journal} {Physical Review Letters}\
  }\textbf {\bibinfo {volume} {125}},\ \bibinfo {pages} {197402} (\bibinfo
  {year} {2020})}\BibitemShut {NoStop}%
\bibitem [{\citenamefont {Lyons}\ \emph {et~al.}(2022)\citenamefont {Lyons},
  \citenamefont {Gillard}, \citenamefont {Leblanc}, \citenamefont {Puebla},
  \citenamefont {Solnyshkov}, \citenamefont {Klompmaker}, \citenamefont
  {Akimov}, \citenamefont {Louca}, \citenamefont {Muduli}, \citenamefont
  {Genco} \emph {et~al.}}]{lyons2022giant}%
  \BibitemOpen
  \bibfield  {author} {\bibinfo {author} {\bibfnamefont {T.}~\bibnamefont
  {Lyons}}, \bibinfo {author} {\bibfnamefont {D.}~\bibnamefont {Gillard}},
  \bibinfo {author} {\bibfnamefont {C.}~\bibnamefont {Leblanc}}, \bibinfo
  {author} {\bibfnamefont {J.}~\bibnamefont {Puebla}}, \bibinfo {author}
  {\bibfnamefont {D.}~\bibnamefont {Solnyshkov}}, \bibinfo {author}
  {\bibfnamefont {L.}~\bibnamefont {Klompmaker}}, \bibinfo {author}
  {\bibfnamefont {I.}~\bibnamefont {Akimov}}, \bibinfo {author} {\bibfnamefont
  {C.}~\bibnamefont {Louca}}, \bibinfo {author} {\bibfnamefont
  {P.}~\bibnamefont {Muduli}}, \bibinfo {author} {\bibfnamefont
  {A.}~\bibnamefont {Genco}}, \emph {et~al.},\ }\bibfield  {title} {\bibinfo
  {title} {Giant effective zeeman splitting in a monolayer semiconductor
  realized by spin-selective strong light--matter coupling},\ }\href
  {https://www.nature.com/articles/s41566-022-01025-8} {\bibfield  {journal}
  {\bibinfo  {journal} {Nature Photonics}\ }\textbf {\bibinfo {volume} {16}},\
  \bibinfo {pages} {632} (\bibinfo {year} {2022})}\BibitemShut {NoStop}%
\bibitem [{\citenamefont {Stier}\ \emph {et~al.}(2016)\citenamefont {Stier},
  \citenamefont {McCreary}, \citenamefont {Jonker}, \citenamefont {Kono},\ and\
  \citenamefont {Crooker}}]{stier2016exciton}%
  \BibitemOpen
  \bibfield  {author} {\bibinfo {author} {\bibfnamefont {A.~V.}\ \bibnamefont
  {Stier}}, \bibinfo {author} {\bibfnamefont {K.~M.}\ \bibnamefont {McCreary}},
  \bibinfo {author} {\bibfnamefont {B.~T.}\ \bibnamefont {Jonker}}, \bibinfo
  {author} {\bibfnamefont {J.}~\bibnamefont {Kono}},\ and\ \bibinfo {author}
  {\bibfnamefont {S.~A.}\ \bibnamefont {Crooker}},\ }\bibfield  {title}
  {\bibinfo {title} {Exciton diamagnetic shifts and valley zeeman effects in
  monolayer ws2 and mos2 to 65 tesla},\ }\href
  {https://www.nature.com/articles/ncomms10643} {\bibfield  {journal} {\bibinfo
   {journal} {Nature communications}\ }\textbf {\bibinfo {volume} {7}},\
  \bibinfo {pages} {10643} (\bibinfo {year} {2016})}\BibitemShut {NoStop}%
\bibitem [{\citenamefont {Stier}\ \emph {et~al.}(2018)\citenamefont {Stier},
  \citenamefont {Wilson}, \citenamefont {Velizhanin}, \citenamefont {Kono},
  \citenamefont {Xu},\ and\ \citenamefont {Crooker}}]{stier2018magnetooptics}%
  \BibitemOpen
  \bibfield  {author} {\bibinfo {author} {\bibfnamefont {A.~V.}\ \bibnamefont
  {Stier}}, \bibinfo {author} {\bibfnamefont {N.~P.}\ \bibnamefont {Wilson}},
  \bibinfo {author} {\bibfnamefont {K.~A.}\ \bibnamefont {Velizhanin}},
  \bibinfo {author} {\bibfnamefont {J.}~\bibnamefont {Kono}}, \bibinfo {author}
  {\bibfnamefont {X.}~\bibnamefont {Xu}},\ and\ \bibinfo {author}
  {\bibfnamefont {S.~A.}\ \bibnamefont {Crooker}},\ }\bibfield  {title}
  {\bibinfo {title} {Magnetooptics of exciton rydberg states in a monolayer
  semiconductor},\ }\href
  {https://journals.aps.org/prl/abstract/10.1103/PhysRevLett.120.057405}
  {\bibfield  {journal} {\bibinfo  {journal} {Physical review letters}\
  }\textbf {\bibinfo {volume} {120}},\ \bibinfo {pages} {057405} (\bibinfo
  {year} {2018})}\BibitemShut {NoStop}%
\bibitem [{\citenamefont {Goryca}\ \emph {et~al.}(2019)\citenamefont {Goryca},
  \citenamefont {Li}, \citenamefont {Stier}, \citenamefont {Taniguchi},
  \citenamefont {Watanabe}, \citenamefont {Courtade}, \citenamefont {Shree},
  \citenamefont {Robert}, \citenamefont {Urbaszek}, \citenamefont {Marie} \emph
  {et~al.}}]{goryca2019revealing}%
  \BibitemOpen
  \bibfield  {author} {\bibinfo {author} {\bibfnamefont {M.}~\bibnamefont
  {Goryca}}, \bibinfo {author} {\bibfnamefont {J.}~\bibnamefont {Li}}, \bibinfo
  {author} {\bibfnamefont {A.~V.}\ \bibnamefont {Stier}}, \bibinfo {author}
  {\bibfnamefont {T.}~\bibnamefont {Taniguchi}}, \bibinfo {author}
  {\bibfnamefont {K.}~\bibnamefont {Watanabe}}, \bibinfo {author}
  {\bibfnamefont {E.}~\bibnamefont {Courtade}}, \bibinfo {author}
  {\bibfnamefont {S.}~\bibnamefont {Shree}}, \bibinfo {author} {\bibfnamefont
  {C.}~\bibnamefont {Robert}}, \bibinfo {author} {\bibfnamefont
  {B.}~\bibnamefont {Urbaszek}}, \bibinfo {author} {\bibfnamefont
  {X.}~\bibnamefont {Marie}}, \emph {et~al.},\ }\bibfield  {title} {\bibinfo
  {title} {Revealing exciton masses and dielectric properties of monolayer
  semiconductors with high magnetic fields},\ }\href
  {https://www.nature.com/articles/s41467-019-12180-y} {\bibfield  {journal}
  {\bibinfo  {journal} {Nature communications}\ }\textbf {\bibinfo {volume}
  {10}},\ \bibinfo {pages} {4172} (\bibinfo {year} {2019})}\BibitemShut
  {NoStop}%
\bibitem [{\citenamefont {Have}\ \emph {et~al.}(2019)\citenamefont {Have},
  \citenamefont {Peres},\ and\ \citenamefont {Pedersen}}]{have2019excitonic}%
  \BibitemOpen
  \bibfield  {author} {\bibinfo {author} {\bibfnamefont {J.}~\bibnamefont
  {Have}}, \bibinfo {author} {\bibfnamefont {N.}~\bibnamefont {Peres}},\ and\
  \bibinfo {author} {\bibfnamefont {T.~G.}\ \bibnamefont {Pedersen}},\
  }\bibfield  {title} {\bibinfo {title} {Excitonic magneto-optics in monolayer
  transition metal dichalcogenides: From nanoribbons to two-dimensional
  response},\ }\href
  {https://journals.aps.org/prb/abstract/10.1103/PhysRevB.100.045411}
  {\bibfield  {journal} {\bibinfo  {journal} {Physical Review B}\ }\textbf
  {\bibinfo {volume} {100}},\ \bibinfo {pages} {045411} (\bibinfo {year}
  {2019})}\BibitemShut {NoStop}%
\bibitem [{\citenamefont {Liu}\ \emph {et~al.}(2019)\citenamefont {Liu},
  \citenamefont {van Baren}, \citenamefont {Taniguchi}, \citenamefont
  {Watanabe}, \citenamefont {Chang},\ and\ \citenamefont
  {Lui}}]{liu2019magnetophotoluminescence}%
  \BibitemOpen
  \bibfield  {author} {\bibinfo {author} {\bibfnamefont {E.}~\bibnamefont
  {Liu}}, \bibinfo {author} {\bibfnamefont {J.}~\bibnamefont {van Baren}},
  \bibinfo {author} {\bibfnamefont {T.}~\bibnamefont {Taniguchi}}, \bibinfo
  {author} {\bibfnamefont {K.}~\bibnamefont {Watanabe}}, \bibinfo {author}
  {\bibfnamefont {Y.-C.}\ \bibnamefont {Chang}},\ and\ \bibinfo {author}
  {\bibfnamefont {C.~H.}\ \bibnamefont {Lui}},\ }\bibfield  {title} {\bibinfo
  {title} {Magnetophotoluminescence of exciton rydberg states in monolayer ws e
  2},\ }\href
  {https://journals.aps.org/prb/abstract/10.1103/PhysRevB.99.205420} {\bibfield
   {journal} {\bibinfo  {journal} {Physical Review B}\ }\textbf {\bibinfo
  {volume} {99}},\ \bibinfo {pages} {205420} (\bibinfo {year}
  {2019})}\BibitemShut {NoStop}%
\bibitem [{\citenamefont {Delhomme}\ \emph {et~al.}(2019)\citenamefont
  {Delhomme}, \citenamefont {Butseraen}, \citenamefont {Zheng}, \citenamefont
  {Marty}, \citenamefont {Bouchiat}, \citenamefont {Molas}, \citenamefont
  {Pan}, \citenamefont {Watanabe}, \citenamefont {Taniguchi}, \citenamefont
  {Ouerghi} \emph {et~al.}}]{delhomme2019magneto}%
  \BibitemOpen
  \bibfield  {author} {\bibinfo {author} {\bibfnamefont {A.}~\bibnamefont
  {Delhomme}}, \bibinfo {author} {\bibfnamefont {G.}~\bibnamefont {Butseraen}},
  \bibinfo {author} {\bibfnamefont {B.}~\bibnamefont {Zheng}}, \bibinfo
  {author} {\bibfnamefont {L.}~\bibnamefont {Marty}}, \bibinfo {author}
  {\bibfnamefont {V.}~\bibnamefont {Bouchiat}}, \bibinfo {author}
  {\bibfnamefont {M.}~\bibnamefont {Molas}}, \bibinfo {author} {\bibfnamefont
  {A.}~\bibnamefont {Pan}}, \bibinfo {author} {\bibfnamefont {K.}~\bibnamefont
  {Watanabe}}, \bibinfo {author} {\bibfnamefont {T.}~\bibnamefont {Taniguchi}},
  \bibinfo {author} {\bibfnamefont {A.}~\bibnamefont {Ouerghi}}, \emph
  {et~al.},\ }\bibfield  {title} {\bibinfo {title} {Magneto-spectroscopy of
  exciton rydberg states in a cvd grown wse2 monolayer},\ }\href
  {https://pubs.aip.org/aip/apl/article-abstract/114/23/232104/37628/Magneto-spectroscopy-of-exciton-Rydberg-states-in}
  {\bibfield  {journal} {\bibinfo  {journal} {Applied Physics Letters}\
  }\textbf {\bibinfo {volume} {114}} (\bibinfo {year} {2019})}\BibitemShut
  {NoStop}%
\bibitem [{\citenamefont {MacNeill}\ \emph {et~al.}(2015)\citenamefont
  {MacNeill}, \citenamefont {Heikes}, \citenamefont {Mak}, \citenamefont
  {Anderson}, \citenamefont {Korm{\'a}nyos}, \citenamefont {Z{\'o}lyomi},
  \citenamefont {Park},\ and\ \citenamefont {Ralph}}]{macneill2015breaking}%
  \BibitemOpen
  \bibfield  {author} {\bibinfo {author} {\bibfnamefont {D.}~\bibnamefont
  {MacNeill}}, \bibinfo {author} {\bibfnamefont {C.}~\bibnamefont {Heikes}},
  \bibinfo {author} {\bibfnamefont {K.~F.}\ \bibnamefont {Mak}}, \bibinfo
  {author} {\bibfnamefont {Z.}~\bibnamefont {Anderson}}, \bibinfo {author}
  {\bibfnamefont {A.}~\bibnamefont {Korm{\'a}nyos}}, \bibinfo {author}
  {\bibfnamefont {V.}~\bibnamefont {Z{\'o}lyomi}}, \bibinfo {author}
  {\bibfnamefont {J.}~\bibnamefont {Park}},\ and\ \bibinfo {author}
  {\bibfnamefont {D.~C.}\ \bibnamefont {Ralph}},\ }\bibfield  {title} {\bibinfo
  {title} {Breaking of valley degeneracy by magnetic field in monolayer mose
  2},\ }\href
  {https://journals.aps.org/prl/abstract/10.1103/PhysRevLett.114.037401}
  {\bibfield  {journal} {\bibinfo  {journal} {Physical review letters}\
  }\textbf {\bibinfo {volume} {114}},\ \bibinfo {pages} {037401} (\bibinfo
  {year} {2015})}\BibitemShut {NoStop}%
\bibitem [{\citenamefont {Klein}\ \emph {et~al.}(2021)\citenamefont {Klein},
  \citenamefont {H{\"o}tger}, \citenamefont {Florian}, \citenamefont
  {Steinhoff}, \citenamefont {Delhomme}, \citenamefont {Taniguchi},
  \citenamefont {Watanabe}, \citenamefont {Jahnke}, \citenamefont {Holleitner},
  \citenamefont {Potemski} \emph {et~al.}}]{klein2021controlling}%
  \BibitemOpen
  \bibfield  {author} {\bibinfo {author} {\bibfnamefont {J.}~\bibnamefont
  {Klein}}, \bibinfo {author} {\bibfnamefont {A.}~\bibnamefont {H{\"o}tger}},
  \bibinfo {author} {\bibfnamefont {M.}~\bibnamefont {Florian}}, \bibinfo
  {author} {\bibfnamefont {A.}~\bibnamefont {Steinhoff}}, \bibinfo {author}
  {\bibfnamefont {A.}~\bibnamefont {Delhomme}}, \bibinfo {author}
  {\bibfnamefont {T.}~\bibnamefont {Taniguchi}}, \bibinfo {author}
  {\bibfnamefont {K.}~\bibnamefont {Watanabe}}, \bibinfo {author}
  {\bibfnamefont {F.}~\bibnamefont {Jahnke}}, \bibinfo {author} {\bibfnamefont
  {A.~W.}\ \bibnamefont {Holleitner}}, \bibinfo {author} {\bibfnamefont
  {M.}~\bibnamefont {Potemski}}, \emph {et~al.},\ }\bibfield  {title} {\bibinfo
  {title} {Controlling exciton many-body states by the electric-field effect in
  monolayer mos 2},\ }\href
  {https://journals.aps.org/prresearch/abstract/10.1103/PhysRevResearch.3.L022009}
  {\bibfield  {journal} {\bibinfo  {journal} {Physical Review Research}\
  }\textbf {\bibinfo {volume} {3}},\ \bibinfo {pages} {L022009} (\bibinfo
  {year} {2021})}\BibitemShut {NoStop}%
\bibitem [{\citenamefont {Van~der Donck}\ \emph {et~al.}(2018)\citenamefont
  {Van~der Donck}, \citenamefont {Zarenia},\ and\ \citenamefont
  {Peeters}}]{PhysRevB.97.195408}%
  \BibitemOpen
  \bibfield  {author} {\bibinfo {author} {\bibfnamefont {M.}~\bibnamefont
  {Van~der Donck}}, \bibinfo {author} {\bibfnamefont {M.}~\bibnamefont
  {Zarenia}},\ and\ \bibinfo {author} {\bibfnamefont {F.~M.}\ \bibnamefont
  {Peeters}},\ }\bibfield  {title} {\bibinfo {title} {Excitons, trions, and
  biexcitons in transition-metal dichalcogenides: Magnetic-field dependence},\
  }\href {https://doi.org/10.1103/PhysRevB.97.195408} {\bibfield  {journal}
  {\bibinfo  {journal} {Phys. Rev. B}\ }\textbf {\bibinfo {volume} {97}},\
  \bibinfo {pages} {195408} (\bibinfo {year} {2018})}\BibitemShut {NoStop}%
\bibitem [{\citenamefont {Varga}(2008)}]{varga2008solution}%
  \BibitemOpen
  \bibfield  {author} {\bibinfo {author} {\bibfnamefont {K.}~\bibnamefont
  {Varga}},\ }\bibfield  {title} {\bibinfo {title} {Solution of few-body
  problems with the stochastic variational method ii: Two-dimensional
  systems},\ }\href
  {https://www.sciencedirect.com/science/article/pii/S0010465508001926}
  {\bibfield  {journal} {\bibinfo  {journal} {Computer Physics Communications}\
  }\textbf {\bibinfo {volume} {179}},\ \bibinfo {pages} {591} (\bibinfo {year}
  {2008})}\BibitemShut {NoStop}%
\bibitem [{\citenamefont {Kidd}\ \emph {et~al.}(2016)\citenamefont {Kidd},
  \citenamefont {Zhang},\ and\ \citenamefont {Varga}}]{kidd2016binding}%
  \BibitemOpen
  \bibfield  {author} {\bibinfo {author} {\bibfnamefont {D.~W.}\ \bibnamefont
  {Kidd}}, \bibinfo {author} {\bibfnamefont {D.~K.}\ \bibnamefont {Zhang}},\
  and\ \bibinfo {author} {\bibfnamefont {K.}~\bibnamefont {Varga}},\ }\bibfield
   {title} {\bibinfo {title} {Binding energies and structures of
  two-dimensional excitonic complexes in transition metal dichalcogenides},\
  }\href {https://journals.aps.org/prb/abstract/10.1103/PhysRevB.93.125423}
  {\bibfield  {journal} {\bibinfo  {journal} {Physical Review B}\ }\textbf
  {\bibinfo {volume} {93}},\ \bibinfo {pages} {125423} (\bibinfo {year}
  {2016})}\BibitemShut {NoStop}%
\bibitem [{\citenamefont {Varga}\ and\ \citenamefont
  {Suzuki}(1997)}]{VARGA1997157}%
  \BibitemOpen
  \bibfield  {author} {\bibinfo {author} {\bibfnamefont {K.}~\bibnamefont
  {Varga}}\ and\ \bibinfo {author} {\bibfnamefont {Y.}~\bibnamefont {Suzuki}},\
  }\bibfield  {title} {\bibinfo {title} {Solution of few-body problems with the
  stochastic variational method i. central forces with zero orbital momentum},\
  }\href@noop {} {\bibfield  {journal} {\bibinfo  {journal} {Computer Physics
  Communications}\ }\textbf {\bibinfo {volume} {106}},\ \bibinfo {pages} {157}
  (\bibinfo {year} {1997})}\BibitemShut {NoStop}%
\bibitem [{\citenamefont {Suzuki}\ and\ \citenamefont
  {Varga}(2014)}]{Suzuki2014-mg}%
  \BibitemOpen
  \bibinfo {editor} {\bibfnamefont {Y.}~\bibnamefont {Suzuki}}\ and\ \bibinfo
  {editor} {\bibfnamefont {K.}~\bibnamefont {Varga}},\ eds.,\ \href@noop {}
  {\emph {\bibinfo {title} {Stochastic variational approach to
  quantum-mechanical few-body problems}}}\ (\bibinfo  {publisher} {Springer},\
  \bibinfo {address} {New York, NY},\ \bibinfo {year} {2014})\BibitemShut
  {NoStop}%
\bibitem [{\citenamefont {Varga}\ and\ \citenamefont
  {Suzuki}(1995)}]{PhysRevC.52.2885}%
  \BibitemOpen
  \bibfield  {author} {\bibinfo {author} {\bibfnamefont {K.}~\bibnamefont
  {Varga}}\ and\ \bibinfo {author} {\bibfnamefont {Y.}~\bibnamefont {Suzuki}},\
  }\bibfield  {title} {\bibinfo {title} {Precise solution of few-body problems
  with the stochastic variational method on a correlated gaussian basis},\
  }\href {https://link.aps.org/doi/10.1103/PhysRevC.52.2885} {\bibfield
  {journal} {\bibinfo  {journal} {Phys. Rev. C}\ }\textbf {\bibinfo {volume}
  {52}},\ \bibinfo {pages} {2885} (\bibinfo {year} {1995})}\BibitemShut
  {NoStop}%
\bibitem [{\citenamefont {Varga}\ \emph {et~al.}(2001)\citenamefont {Varga},
  \citenamefont {Navratil}, \citenamefont {Usukura},\ and\ \citenamefont
  {Suzuki}}]{PhysRevB.63.205308}%
  \BibitemOpen
  \bibfield  {author} {\bibinfo {author} {\bibfnamefont {K.}~\bibnamefont
  {Varga}}, \bibinfo {author} {\bibfnamefont {P.}~\bibnamefont {Navratil}},
  \bibinfo {author} {\bibfnamefont {J.}~\bibnamefont {Usukura}},\ and\ \bibinfo
  {author} {\bibfnamefont {Y.}~\bibnamefont {Suzuki}},\ }\bibfield  {title}
  {\bibinfo {title} {Stochastic variational approach to few-electron artificial
  atoms},\ }\href {https://doi.org/10.1103/PhysRevB.63.205308} {\bibfield
  {journal} {\bibinfo  {journal} {Phys. Rev. B}\ }\textbf {\bibinfo {volume}
  {63}},\ \bibinfo {pages} {205308} (\bibinfo {year} {2001})}\BibitemShut
  {NoStop}%
\bibitem [{\citenamefont {Yan}\ and\ \citenamefont
  {Varga}(2020)}]{PhysRevB.101.235435}%
  \BibitemOpen
  \bibfield  {author} {\bibinfo {author} {\bibfnamefont {J.}~\bibnamefont
  {Yan}}\ and\ \bibinfo {author} {\bibfnamefont {K.}~\bibnamefont {Varga}},\
  }\bibfield  {title} {\bibinfo {title} {Excited-state trions in
  two-dimensional materials},\ }\href
  {https://doi.org/10.1103/PhysRevB.101.235435} {\bibfield  {journal} {\bibinfo
   {journal} {Phys. Rev. B}\ }\textbf {\bibinfo {volume} {101}},\ \bibinfo
  {pages} {235435} (\bibinfo {year} {2020})}\BibitemShut {NoStop}%
\bibitem [{\citenamefont {Lundt}\ \emph {et~al.}(2016)\citenamefont {Lundt},
  \citenamefont {Klembt}, \citenamefont {Cherotchenko}, \citenamefont
  {Betzold}, \citenamefont {Iff}, \citenamefont {Nalitov}, \citenamefont
  {Klaas}, \citenamefont {Dietrich}, \citenamefont {Kavokin}, \citenamefont
  {H{\"o}fling} \emph {et~al.}}]{lundt2016room}%
  \BibitemOpen
  \bibfield  {author} {\bibinfo {author} {\bibfnamefont {N.}~\bibnamefont
  {Lundt}}, \bibinfo {author} {\bibfnamefont {S.}~\bibnamefont {Klembt}},
  \bibinfo {author} {\bibfnamefont {E.}~\bibnamefont {Cherotchenko}}, \bibinfo
  {author} {\bibfnamefont {S.}~\bibnamefont {Betzold}}, \bibinfo {author}
  {\bibfnamefont {O.}~\bibnamefont {Iff}}, \bibinfo {author} {\bibfnamefont
  {A.~V.}\ \bibnamefont {Nalitov}}, \bibinfo {author} {\bibfnamefont
  {M.}~\bibnamefont {Klaas}}, \bibinfo {author} {\bibfnamefont {C.~P.}\
  \bibnamefont {Dietrich}}, \bibinfo {author} {\bibfnamefont {A.~V.}\
  \bibnamefont {Kavokin}}, \bibinfo {author} {\bibfnamefont {S.}~\bibnamefont
  {H{\"o}fling}}, \emph {et~al.},\ }\bibfield  {title} {\bibinfo {title}
  {Room-temperature tamm-plasmon exciton-polaritons with a wse2 monolayer},\
  }\href {https://www.nature.com/articles/ncomms13328} {\bibfield  {journal}
  {\bibinfo  {journal} {Nature communications}\ }\textbf {\bibinfo {volume}
  {7}},\ \bibinfo {pages} {13328} (\bibinfo {year} {2016})}\BibitemShut
  {NoStop}%
\bibitem [{\citenamefont {Gu}\ \emph {et~al.}(2019)\citenamefont {Gu},
  \citenamefont {Chakraborty}, \citenamefont {Khatoniar},\ and\ \citenamefont
  {Menon}}]{gu2019room}%
  \BibitemOpen
  \bibfield  {author} {\bibinfo {author} {\bibfnamefont {J.}~\bibnamefont
  {Gu}}, \bibinfo {author} {\bibfnamefont {B.}~\bibnamefont {Chakraborty}},
  \bibinfo {author} {\bibfnamefont {M.}~\bibnamefont {Khatoniar}},\ and\
  \bibinfo {author} {\bibfnamefont {V.~M.}\ \bibnamefont {Menon}},\ }\bibfield
  {title} {\bibinfo {title} {A room-temperature polariton light-emitting diode
  based on monolayer ws2},\ }\href
  {https://www.nature.com/articles/s41565-019-0543-6} {\bibfield  {journal}
  {\bibinfo  {journal} {Nature nanotechnology}\ }\textbf {\bibinfo {volume}
  {14}},\ \bibinfo {pages} {1024} (\bibinfo {year} {2019})}\BibitemShut
  {NoStop}%
\bibitem [{\citenamefont {Anton-Solanas}\ \emph {et~al.}(2021)\citenamefont
  {Anton-Solanas}, \citenamefont {Waldherr}, \citenamefont {Klaas},
  \citenamefont {Suchomel}, \citenamefont {Harder}, \citenamefont {Cai},
  \citenamefont {Sedov}, \citenamefont {Klembt}, \citenamefont {Kavokin},
  \citenamefont {Tongay} \emph {et~al.}}]{anton2021bosonic}%
  \BibitemOpen
  \bibfield  {author} {\bibinfo {author} {\bibfnamefont {C.}~\bibnamefont
  {Anton-Solanas}}, \bibinfo {author} {\bibfnamefont {M.}~\bibnamefont
  {Waldherr}}, \bibinfo {author} {\bibfnamefont {M.}~\bibnamefont {Klaas}},
  \bibinfo {author} {\bibfnamefont {H.}~\bibnamefont {Suchomel}}, \bibinfo
  {author} {\bibfnamefont {T.~H.}\ \bibnamefont {Harder}}, \bibinfo {author}
  {\bibfnamefont {H.}~\bibnamefont {Cai}}, \bibinfo {author} {\bibfnamefont
  {E.}~\bibnamefont {Sedov}}, \bibinfo {author} {\bibfnamefont
  {S.}~\bibnamefont {Klembt}}, \bibinfo {author} {\bibfnamefont {A.~V.}\
  \bibnamefont {Kavokin}}, \bibinfo {author} {\bibfnamefont {S.}~\bibnamefont
  {Tongay}}, \emph {et~al.},\ }\bibfield  {title} {\bibinfo {title} {Bosonic
  condensation of exciton--polaritons in an atomically thin crystal},\ }\href
  {https://www.nature.com/articles/s41563-021-01000-8} {\bibfield  {journal}
  {\bibinfo  {journal} {Nature materials}\ }\textbf {\bibinfo {volume} {20}},\
  \bibinfo {pages} {1233} (\bibinfo {year} {2021})}\BibitemShut {NoStop}%
\bibitem [{\citenamefont {Claudio~Andreani}(1995)}]{claudio1995optical}%
  \BibitemOpen
  \bibfield  {author} {\bibinfo {author} {\bibfnamefont {L.}~\bibnamefont
  {Claudio~Andreani}},\ }\bibfield  {title} {\bibinfo {title} {Optical
  transitions, excitons, and polaritons in bulk and low-dimensional
  semiconductor structures},\ }in\ \href@noop {} {\emph {\bibinfo {booktitle}
  {Confined Electrons and photons: New physics and applications}}}\ (\bibinfo
  {publisher} {Springer},\ \bibinfo {year} {1995})\ pp.\ \bibinfo {pages}
  {57--112}\BibitemShut {NoStop}%
\bibitem [{\citenamefont {Gor'kov}\ and\ \citenamefont
  {Dzyaloshinskii}(1968)}]{Gorkov1967}%
  \BibitemOpen
  \bibfield  {author} {\bibinfo {author} {\bibfnamefont {L.~P.}\ \bibnamefont
  {Gor'kov}}\ and\ \bibinfo {author} {\bibfnamefont {I.~E.}\ \bibnamefont
  {Dzyaloshinskii}},\ }\bibfield  {title} {\bibinfo {title} {Contribution to
  the theory of the mott exciton in a strong magnetic field},\ }\href
  {http://www.jetp.ras.ru/cgi-bin/e/index/r/53/2/p717?a=list} {\bibfield
  {journal} {\bibinfo  {journal} {JETP}\ }\textbf {\bibinfo {volume} {26}},\
  \bibinfo {pages} {449} (\bibinfo {year} {1968})}\BibitemShut {NoStop}%
\bibitem [{\citenamefont {Na}\ and\ \citenamefont
  {Rhee}(2006)}]{na2006electronic}%
  \BibitemOpen
  \bibfield  {author} {\bibinfo {author} {\bibfnamefont {M.}~\bibnamefont
  {Na}}\ and\ \bibinfo {author} {\bibfnamefont {S.-W.}\ \bibnamefont {Rhee}},\
  }\bibfield  {title} {\bibinfo {title} {Electronic characterization of al/pmma
  [poly (methyl methacrylate)]/p-si and al/cep (cyanoethyl pullulan)/p-si
  structures},\ }\href
  {https://www.sciencedirect.com/science/article/pii/S1566119906000267}
  {\bibfield  {journal} {\bibinfo  {journal} {Organic electronics}\ }\textbf
  {\bibinfo {volume} {7}},\ \bibinfo {pages} {205} (\bibinfo {year}
  {2006})}\BibitemShut {NoStop}%
\bibitem [{\citenamefont {Khestanova}\ \emph {et~al.}(2024)\citenamefont
  {Khestanova}, \citenamefont {Shahnazaryan}, \citenamefont {Kozin},
  \citenamefont {Kondratyev}, \citenamefont {Krizhanovskii}, \citenamefont
  {Skolnick}, \citenamefont {Shelykh}, \citenamefont {Iorsh},\ and\
  \citenamefont {Kravtsov}}]{khestanova2024electrostatic}%
  \BibitemOpen
  \bibfield  {author} {\bibinfo {author} {\bibfnamefont {E.}~\bibnamefont
  {Khestanova}}, \bibinfo {author} {\bibfnamefont {V.}~\bibnamefont
  {Shahnazaryan}}, \bibinfo {author} {\bibfnamefont {V.~K.}\ \bibnamefont
  {Kozin}}, \bibinfo {author} {\bibfnamefont {V.~I.}\ \bibnamefont
  {Kondratyev}}, \bibinfo {author} {\bibfnamefont {D.~N.}\ \bibnamefont
  {Krizhanovskii}}, \bibinfo {author} {\bibfnamefont {M.~S.}\ \bibnamefont
  {Skolnick}}, \bibinfo {author} {\bibfnamefont {I.~A.}\ \bibnamefont
  {Shelykh}}, \bibinfo {author} {\bibfnamefont {I.~V.}\ \bibnamefont {Iorsh}},\
  and\ \bibinfo {author} {\bibfnamefont {V.}~\bibnamefont {Kravtsov}},\
  }\bibfield  {title} {\bibinfo {title} {Electrostatic control of nonlinear
  photonic-crystal polaritons in a monolayer semiconductor},\ }\href
  {https://arxiv.org/abs/2402.16193} {\bibfield  {journal} {\bibinfo  {journal}
  {arXiv:2402.16193}\ } (\bibinfo {year} {2024})}\BibitemShut {NoStop}%
\end{thebibliography}
\end{document}